\documentclass[letterpaper,12pt]{article} 

\usepackage{amsmath}
\usepackage{amssymb}
\usepackage[normalem]{ulem}
\usepackage[]{natbib}
\usepackage[dvips]{epsfig}
\usepackage{dcolumn}
\usepackage{enumerate}
\usepackage{hhline}
\usepackage{dsfont}
\usepackage{afterpage}
\usepackage{arydshln}
\usepackage{graphicx}
\usepackage{color}

\usepackage[usenames,dvipsnames]{xcolor}
\usepackage{rotating}
\PassOptionsToPackage{hyphens}{url}
\usepackage[breaklinks]{hyperref}
\usepackage{breakurl} 
\usepackage{xr}
\usepackage[percent]{overpic}
\usepackage{subfig}
\usepackage[]{caption}
\usepackage{algorithmicx}
\usepackage[noend]{algpseudocode}
\usepackage{algorithm}
\usepackage{diagbox}
\usepackage{graphicx}
\usepackage{wrapfig}
\usepackage{lscape}
\input epsf
\usepackage{fontenc}
\usepackage{setspace}
\usepackage{bm}
\usepackage{slashbox}

\usepackage{multirow}
\usepackage{eurosym}
\usepackage{titlesec}

\epsfverbosetrue
\def\theequation{\thesection.\arabic{equation}}  
\def\abstract{\if@twocolumn
\section*{Abstract}
\else \normalsize 
\begin{center}
{\bf Summary\vspace{-.5em}\vspace{0pt}} 
\end{center}
\quotation 
\fi}
\def\endabstract{\if@twocolumn\else\endquotation\fi}

\makeatletter
\newcommand{\myappendix}[1]{
	\setcounter{section}{1}
        \renewcommand{\thesection}{A\arabic{section}}}

\DeclareMathAlphabet{\mathpzc}{OT1}{pzc}{m}{it}
\DeclareFontFamily{OT1}{pzc}{}
\DeclareFontShape{OT1}{pzc}{m}{it}{<-> s * [1.200] pzcmi7t}{}

\newcommand{\supnorm}[1]{\left\lVert#1\right\rVert_{\infty}}

\def \dsP {\text{$\mathds{P}$}}
\def \dsE {\text{$\mathds{E}$}}
\def \dsR {\text{$\mathds{R}$}}

\DeclareMathOperator{\rank}{rk}

\DeclareMathOperator{\pen}{\scriptsize{penalized}}
\DeclareMathOperator{\unpen}{\scriptsize{unpenalized}}

\DeclareMathOperator{\spa}{span}

\DeclareMathOperator{\ND}{\mathcal{N}}

\DeclareMathOperator{\GaD}{\mathpzc{Ga}}

\DeclareMathOperator{\ExpD}{\mathpzc{Exp}}

\DeclareMathOperator{\BerD}{\mathpzc{Be}}

\DeclareMathOperator{\IGD}{\mathcal{IG}}
 \newcommand{\BetaD}{\mathpzc{Beta}}
 \DeclareMathOperator{\invGD}{\mathpzc{InvGauss}}

\def \calD {\mathcal D}

    \def \mA {\text{\boldmath$A$}}
\def \bvec {\text{\boldmath$b$}}    \def \mB {\text{\boldmath$B$}}
    
    \def \mD {\text{\boldmath$D$}}
    
\def \fvec {\text{\boldmath$f$}}

    \def \mI {\text{\boldmath$I$}}
    
    \def \mK {\text{\boldmath$K$}}

\def \wvec {\text{\boldmath$w$}}    
\def \xvec {\text{\boldmath$x$}}    \def \mX {\text{\boldmath$X$}}
\def \yvec {\text{\boldmath$y$}}    
    
\def \nuvec {\text{\boldmath$\nu$}}

\def \betavec         {\text{\boldmath$\beta$}}

\def \etavec          {\text{\boldmath$\eta$}}
\def \thetavec        {\text{\boldmath$\theta$}}

\def \muvec           {\text{\boldmath$\mu$}}
\def \nuvec           {\text{\boldmath$\nu$}}

\def \betatildevec         {\text{\boldmath$\tilde \beta$}}

\def \mSigma   {\mathbf{\Sigma}}

\def \nullvec {\mathbf{0}}

\usepackage{color}
\usepackage{colordvi}
\newlength{\breite}
\breite\textwidth
\newcounter{aufg}[section]
  {\refstepcounter{aufg}\noindent\textbf{Exercise \arabic{aufg}:}
   \\*[1ex]\noindent}{\vspace{.5cm}}
   
 \newcounter{notes}[section]
  {\refstepcounter{aufg}\noindent\textbf{}
   \\*[1ex]\noindent}{\vspace{.5cm}}
   
\usepackage{amsthm}  

\theoremstyle{definition}

\newtheorem*{beisp*}{Example}
\newtheorem{Proof}{Proof}


\newtheoremstyle{break}
  {}
  {}
  {}
  {}
  {\bfseries}
  {.}
  {\newline}
  {}
  
\theoremstyle{break}

\newcommand{\head}[2]%
 {\hrule \vspace{.15cm} {\sfbold Advanced Statistical Inference, Summer Term 2012, Georg-August-University G\"ottingen}\hfill
{\sfbold Sheet #1}\\
{\sfbold Prof. Dr. Thomas Kneib, Nadja Klein}\hfill {\sfbold #2}

\vspace{.2cm}
\hrule

\vspace{1cm}

}

\newcounter{auf}
{\refstepcounter{auf}
\begin{center}
\fcolorbox[gray]{0}{.95}{
\makebox[\breite]{
\textbf{Exercise \arabic{auf}}
}}\\*[1ex]\noindent
\end{center}
}{\vspace{.5cm}}

\newcounter{loes}[section]
{\stepcounter{loes}
\begin{center}
\fcolorbox[gray]{0}{.95}{
\makebox[\breite]{
\textbf{L"osung \arabic{loes}}
}}\\*[1ex]\noindent
\end{center}
}{}

{\begin{center}
\fcolorbox[gray]{0}{.95}{
\makebox[\breite]{
\textbf{Zu Aufgabe #1}
}}\\*[1ex]\noindent
\end{center}\vspace{1cm}
}{\vspace{1cm}}

\newcounter{ka}
{\refstepcounter{ka}
\begin{center}
\framebox[\textwidth]{
\textbf{Aufgabe \arabic{ka}} \hfill #1 Punkte
}\\*[1ex]\noindent
\end{center}
}{\vspace{1cm}}

\newcounter{lka}
{\refstepcounter{lka}
\begin{center}
\framebox[\textwidth]{
\textbf{L\"osung \arabic{lka}} \hfill #1 Punkte
}\\*[1ex]\noindent
\end{center}
}{\vspace{1cm}}

\usepackage[margin=1in]{geometry}
\titlespacing*\section{0pt}{5pt plus 4pt minus 2pt}{5pt plus 2pt minus 2pt}
\titlespacing*\subsection{0pt}{5pt plus 4pt minus 2pt}{5pt plus 2pt minus 2pt}
\titlespacing*\subsubsection{0pt}{0pt plus 4pt minus 2pt}{0pt plus 2pt minus 2pt}
\titlespacing*\paragraph{0pt}{5pt plus 4pt minus 2pt}{8pt plus 2pt minus 2pt}

\definecolor{parisgreen}{rgb}{0.31, 0.78, 0.47}
\definecolor{pinegreen}{rgb}{0.0, 0.47, 0.44}
\definecolor{midnightblue}{rgb}{0.1, 0.1, 0.44}

\newcounter{myremark}

\newcommand{\ns}[1]{{\leavevmode\color{cyan}\sout{#1}}}

\newcounter{mynotation}

\usepackage{paralist}

\makeatletter
\def\@seccntformat#1{\@ifundefined{#1@cntformat}%
	{\csname the#1\endcsname\quad}  
	{\csname #1@cntformat\endcsname}
}
\let\oldappendix\appendix 
\renewcommand\appendix{%
	\oldappendix
	\newcommand{\section@cntformat}{\appendixname~\thesection\quad}
}
\makeatother

\renewcommand{\baselinestretch}{1.4}

\begin{document}
\setlength{\abovedisplayskip}{0.15cm}
\setlength{\belowdisplayskip}{0.15cm}
\pagestyle{empty}
\begin{titlepage}

\title{Bayesian Effect Selection for Additive Quantile Regression  with an Analysis to Air Pollution Thresholds}

\author{Nadja Klein 
 and Jorge Mateu 
}
\date{}
\maketitle
\noindent
\vfill
{\small Nadja Klein is Assistant Professor of Applied Statistics and Emmy Noether Research Group Leader in Statistics and Data Science at Humboldt-Universit\"at zu Berlin; Jorge Mateu is Professor of Statistics within the Department of Mathematics at Universitat Jaume I. Correspondence should be directed to~Prof.~Dr.~Nadja Klein at Humboldt Universit\"at zu Berlin,
Unter den Linden 6, 10099 Berlin. Email: nadja.klein@hu-berlin.de.

\vfill
\noindent \textbf{Acknowledgments:} Nadja Klein gratefully acknowledges support  by the German research foundation (DFG) through the  Emmy Noether grant KL 3037/1-1. {Jorge Mateu is partially supported by 
grants PID2019-107392RB-I00 from Ministry of Science and Innovation, and by UJI-B2018-04 from Universitat Jaume I.}}\\

\newpage
\begin{center}
\mbox{}\vspace{2cm}\\
{\LARGE \title{Bayesian Effect Selection for Additive Quantile Regression  with an Analysis to Air Pollution Thresholds}
}\\
\vspace{1.5cm}
{\Large Abstract}
\end{center}
\vspace{-1pt}
\onehalfspacing
\noindent

Statistical techniques used in air pollution modelling usually lack the possibility to understand which predictors affect air pollution in which functional form; and are not able to regress on exceedances over certain thresholds imposed by authorities directly. The latter naturally induce conditional quantiles and reflect the seriousness of particular events. In the present paper we focus on this important aspect by developing quantile regression models further.  We propose a general Bayesian effect selection approach for additive quantile regression within a highly interpretable framework. We place separate normal beta prime spike and slab priors on the scalar importance parameters of effect parts and implement a fast Gibbs sampling scheme. Specifically, it enables to study quantile-specific covariate effects, allows these covariates to be of general functional form using additive predictors, and facilitates the analysts' decision whether an effect should be included linearly, non-linearly or not at all in the quantiles of interest.  In a detailed analysis on air pollution data in Madrid (Spain) we find the added value of modelling extreme nitrogen dioxide (NO2) concentrations and how thresholds are driven differently by several climatological variables and traffic as a spatial proxy.  Our results underpin the need of enhanced statistical models to support short-term decisions and enable local authorities to mitigate or even prevent exceedances of NO2 concentration limits.

\vspace{20pt}
 
\noindent
{\bf Keywords}: Air pollution; asymmetric Laplace distribution; distributional regression; effect decomposition; NO2 and O3; penalised splines; spike and slab prior 
\end{titlepage}

\newpage
\pagestyle{plain}
\setcounter{equation}{0}
\renewcommand{\theequation}{\arabic{equation}}

\section{Introduction}\label{sec:intro}

Air quality is deteriorating globally at an alarming rate due to increasing industrialisation
and urbanisation~\citep{Ryu2019}. According to the European Environment Agency \citep{EEA}, air pollution is the single largest environmental health risk in Europe, and in many countries worldwide.
An expert committee, being part of the British Committee on the Medical Effects of Air Pollutants, estimated in August 2018 that between 28,000 and 36,000 premature deaths may be linked to air pollution in the UK every year \citep{airquality}. This report complies with the study presented in 2018 by European Environment Agency~\citep{Eur2018}, which states that the estimated impacts on the population in 41 European countries of exposure to nitrogen dioxide (NO2) and ozone (O3) concentrations in 2015 were
around 79,000 and 17,700 premature deaths per year,
respectively.

Many European cities still regularly exceed current EU limits for NO2~\citep{Eur2018}. Indeed, the Urban NO2 Atlas \citep{NO2Atlas} provides city {fact-sheets} to help designing effective air quality measures {with the aim to} reduce NO2 concentration within European cities.
The Atlas identifies the main sources of NO2 pollution for each city, {helping policymakers to design target measures and actions against them.}

{NO2} is one
of a group of gases called nitrogen oxides (NOx). While all of these gases are harmful to human health and the environment, NO2 is of greater concern~\citep{Val2011,Ach2019} {and is mostly emitted into the air through} the burning of fuel. NO2 generates from emissions from cars, trucks and buses, power plants, and off-road equipment~\citep{Per2001,Lee2014,Cat2016} {and is a primary pollutant}. {Indeed, as indicated by the European Environment Agency~\citep{Eur2018}, road transport is the largest contributor to NO2 pollution in the EU, ahead of the energy, commercial, institutional and household sectors}. On the other side, the ozone molecule{s} (O3) {are} harmful to air quality outside of the ozone layer. {O3 is a so-called secondary pollutant not emitted directly from a source (like vehicles or power plants).}
O3 can be ``good'' or ``bad'' for health and the environment depending on where it is found in the atmosphere. Stratospheric ozone is ``good'' because it protects {the living organisms from} ultraviolet radiation from the sun. Ground-level ozone {is a colorless and highly irritating gas that forms just above the earth's surface. It is ``bad''} because it can trigger a variety of {human} health problems, particularly for children, the elderly, and people of all ages who have lung diseases such as asthma~\citep{Val2011}.
 Breathing air with a high concentration of NO2 can irritate airways in the human respiratory system and can harm our health {as well}.
As the global population becomes more health conscious, various studies have been conducted to determine the effect of NO2 concentration on human health. High NO2 concentrations in urban areas cause bronchial and lung cancer and have severe effects on asthmatic patients~\citep{Ach2019}.

Social medical costs due to NO2 pollution are certainly high. To reduce these social costs, many countries regulate NO2 concentration levels using environmental
policies {targeting at} NO2 reduction~\citep{Val2011,Bor2018,Ryu2019}. For example, the EU has established an integrated environmental
policy agreement for the transport, industrial, and energy sectors to improve air pollution at national,
regional, and local levels. In addition, since 2013, China has sought to install selective catalytic
reduction (SCR) equipment in power plants to establish emissions standards to reduce NO2 levels
through the Air Pollution Prevention and Control Action Plan.

However, to establish effective environmental regulations that reduce the impact of NO2, accurate information on the nature and the way several sources of air pollution interact with and/or produce NO2 is by {all} means needed and essential. 
Modelling air pollution at both global and local scales enables consistent comparisons of relations between air pollution and health. For example, as NO2 is highly traffic-related and {localised}, its local understanding is needed for the assessment of personal outdoor exposure, in particular in areas close to sources of the pollutant such as primary roads.

Statistical models for air pollution analysis are under continuous development. Most of the classical approaches are based on regression techniques through linear-based relations~\citep{Men2015,Hat2017,Ant2018,Lua2020}. {In addition,} ensemble tree-based {(e.g.~random forests)} and neural network techniques~\citep{Ryu2019,Pry2000} {have been considered} to investigate if more flexible models can better capture non-linear relationships between predictors and NO2.
Regression-based methods fit one model to the
entire range of each predictor, while ensemble tree-based methods build on subsets of data and sub-ranges of predictors. {The latter methods}  are representative for the techniques that are evaluated in the most recent air pollution modelling.
{However, despite the increased flexibility such models lack the possibility to understand and interpret which predictors affect air pollution and in what functional form (e.g.~linearly or non-linearly). Both approaches in addition come with the drawback of modelling  the expected NO2 concentration only rather than allowing to directly regress on exceedances over certain thresholds  of the entire NO2 distributions.  }
{Such thresholds are imposed by authorities} to decide on and classify the seriousness of a particular event.  { From a statistical point of view, these thresholds induce  quantiles,} and we focus here on this interesting (while not much treated) aspect by {using} quantile regression {models}. {However, in addition to thereby allowing the predictor effects to differ for distinct quantiles, we develop quantile regression further and facilitate general function selection and effect decomposition (e.g.~into respective linear and non-linear parts) in a highly interpretable model.}

{To determine the influence of covariates $\mX=(X_1,\ldots,X_p)^\top$ on quantiles 
of the distribution of a dependent variable $Y\in\mathcal{Y}\subset\dsR$  directly, an important contribution amongst other functionals beyond the mean~\citep{Kne2012} is quantile regression~\citep{Koe2005}.  One of the main
advantages over mean regression~\citep{McCNel1989,HasTib1990} is that quantile regression permits to supply detailed
information about the complete conditional distribution instead of only the mean by considering a dense set of conditional quantiles. In
addition, outliers and extreme data are usually less influential in quantile regression
due to the inherent robustness of quantiles.

Estimation of the conditional $\tau$th quantile of interest, $\tau\in(0,1)$ in a basic linear quantile regression (QR) model, as proposed in~\citet{KoeBas1978}, is based on solving 
\begin{equation}\label{eq:awad}
    \min_{\beta\in\dsR^p}\sum_{i=1}^n \rho_\tau(y_i-\xvec_i\betavec),
\end{equation}
where $\rho_{\tau}(u)=u(\tau-\mathds{1}_{\lbrace u<0\rbrace})$ for $u\in\dsR$ is the so-called piecewise linear ``check function''. This approach is completely non-parametric as it does not make any specific choice about an error distribution. No closed form solution for the minimisation of the above problem exists, but quantile regression (QR) estimates can be obtained
based on linear programming.}

{To be able to {handle} non-linear or more general functional relationships between response and covariates on conditional quantiles directly, structured additive quantile (STAQ) regression models have been suggested in the literature~\citep{YueRue2011,WalKneLanYue2012}. These build on the general idea of structured additive models for mean regression~\citep[STAR;][]{FahKneLan2004,Woo2017} which can capture not only linear and non-linear shapes of univariate covariates but also spatial effects, interactions, grouping effects and others. While the aforementioned references to STAQ models rely on Bayesian principles, alternatives have been developed as well \citep[see e.g.][]{Mei2006,Koe2010,FenKneHot2011,Koe2011,FasWooZafNedGou2020}.}

{However, {when} it comes to variable or effect selection in such STAQ {models, literature,} becomes much scarcer. Indeed, variable selection for quantile regression has so far mostly been done in the linear case, see for instance \cite{WuLiu2009,Alh2015} for a few references within a classical frequentist  framework and \cite{AlhYu2012,AlhYu2013,YuCheReeDun2013} for some Bayesian counterparts.}

{To fill this  gap and motivated by our application to air pollution, it is thus the aim of this paper to  develop general Bayesian effect selection for STAQ models thereby significantly enriching existing suggestions for linear quantile regression. To achieve this goal, we employ a normal beta prime spike and slab prior on the scalar importance parameters of predictor effect parts~\citet{KleCarKneLanWag2021} and implement posterior estimation based on efficient Markov chain Monte Carlo (MCMC) simulations.  All steps can be {realised} by Gibbs updates thus being fast to draw samples from the complex posteriors. 

Bayesian variable selection based on spike and slab priors has been {popularised} by~\citet{MitBea1988} for linear mean regression and was significantly developed further in several contexts~\citep[see, for instance,][]{GeoMcC1997}; see also \citet{ClyGeo2004,OHaSil2009} for some overviews. We believe our  approach makes a necessary contribution and specifically enables for three essential features simultaneously: to (i) study  quantile-specific  covariate effects, (ii) allow these covariates to be of general functional form (e.g.~non-linear) using additive predictors, (iii) decide whether an effect should be included linearly, non-linearly or not at all in the  relevant threshold quantiles. In our application the latter will be  $\tau\in\lbrace 0.6,0.8,0.9\rbrace$, see  Section~\ref{sec:data} for more details.}

{The rest of the paper is structured as follows. Inspired by our application and analysis with background and data treated in Section~\ref{sec:data}, we detail our methodological approach to Bayesian effect selection in STAQ models
 in Section~\ref{sec:effselQR}. 
While our method can be applied in a wide range of applications, our empirical results in Section~\ref{sec:results} quantify the added-value of modelling extreme NO2 concentrations, and how thresholds inducing conditional quantiles are able to highlight important modelling differences. These results underpin the need of
enhanced statistical models to support short-term decisions and enable local authorities to mitigate or even prevent exceedances of NO2 concentration limits. {The paper ends with some conclusions and a discussion in Section~\ref{sec:conclusion}}.}

\section{The data and research questions}\label{sec:data}

\subsection{Air pollution, NO2 and O3. Some motivating basics}\label{subsec:motivation}

Air pollution in urban areas is mainly due to the intense use of motorised transport for travelling, in particular private cars and heavy goods vehicles. This is a priority issue for transportation planners and public authorities, given the harmful effects of pollution to human health and the environment~\citep{Berg2013}. 
Nitrogen dioxide (NO2) along with  nitric oxide (NO) reacts with other chemicals in the air to form both particulate matter and ozone. Both of these are also harmful when inhaled due to effects on the respiratory system. NO2 and NO interact with water, oxygen and other chemicals in the atmosphere to form acid rain. Acid rain harms sensitive ecosystems such as lakes and forests. Breathing air with a high concentration of NO2 can irritate airways in the human respiratory system. 
Such exposures over short periods can aggravate respiratory diseases, particularly asthma, leading to respiratory symptoms (such as coughing, wheezing or difficulty breathing), hospital admissions and visits to emergency rooms. Longer exposures to elevated concentrations of NO2 may contribute to the development of asthma and potentially increase susceptibility to respiratory infections. People with asthma, as well as children and the elderly, are generally at greater risk for  the health effects of NO2.

Ground{-}level ozone  (O3)  is not emitted directly into the air, but is created by chemical reactions between oxides of nitrogen (NOx) and volatile organic compounds (VOC). This happens when pollutants emitted by cars, power plants, industrial boilers, refineries, chemical plants, and other sources chemically react in the presence of sunlight. Ozone in the air we breathe can harm our health. 
People with certain genetic characteristics, and people with reduced intake of certain nutrients, such as vitamins C and E, are at greater risk from ozone exposure. Breathing elevated concentrations of O3 can trigger a variety of responses, such as chest pain, coughing, throat irritation, and airway inflammation. It also can reduce lung function and harm lung tissue. Ozone can worsen bronchitis, emphysema, and asthma, leading to increased medical care.

One of the main problems caused by air pollution in urban areas is photochemical oxidants. Among these, {O3} and NO2 are particularly important because they are capable of causing adverse effects on human health~\citep{Who2014}. The formation of {O3} depends on the intensity of solar radiation, the
absolute concentrations of NOx and VOC, and the ratio of NOx (NO and NO2) to VOC~\citep{Val2011}. NO is converted to NO2 via a reaction with O3, and during daytime hours NO2 is converted back to NO as a result of photolysis, which leads to the regeneration of O3. However, this phenomenon is not well understood so far.
The concentration of photochemical oxidants can be
decreased by controlling their precursors: nitrogen oxides NOx  and VOCs. However, the efficiency of emission control also depends on the relationship between primary and
secondary pollutants, as well as ambient meteorological conditions. 

Clearly, and owing to the chemical coupling of O3 and NOx, the levels of O3 and NO2 are inherently linked. Therefore, the response to reduction in the emission of NOx is
remarkably non-linear~\citep{Val2011,Bor2018,Ryu2019} and any resultant reduction in the level of NO2 is invariably accompanied by an increase in the level of O3. In addition, changes in the level of O3 on a global scale lead to an increasing background which influences local O3 and NO2 levels and the effectiveness of local emission controls. 

It is therefore necessary {to obtain a thorough understanding of the cross relationship among O3, traffic flow and NO2 under various atmospheric conditions to improve the understanding of the chemical coupling among them}. The non-linear mechanisms of the inter-dependencies between O3 and NO2 in combination with other pollutants and atmospheric conditions are still not well understood and need further study. These non-linear effects are in line with thresholds that authorities impose on the records of NO2; environmental pollution alarms are placed upon three different thresholds, 60\% (moderate), 80\% (large) and 90\% (extreme).

\subsection{{The data}}\label{subsec:datades}

The Surveillance System of the City of Madrid (Spain) keeps a web portal with open data from many sources and environmental problems (see \burl{https://datos.madrid.es/portal/site/egob}). In particular, this system collects through a number of stations basic information for atmospheric
surveillance. We focus {here} on one weather and one pollution station located in downtown Madrid as they are {placed} in such a particular place of the city that represents one of the peak locations of air pollution and traffic congestion. 
We have daily data from 1 January 2016 to 31 December 2019 on the three air pollutants NO2 ($\mathit{no2}$), O3 ($\mathit{o3}$) and CO (carbon monoxide, $\mathit{co}$), {along with a number of climatological variables}. For the latter we have daily average precipitation ($\mathit{prec}$), temperature ($\mathit{temp}$), average wind speed {($\mathit{vel}$)}, wind gust speed {($\mathit{racha}$)}, maximum pressure {($\mathit{pres\_max}$)}, and minimum pressure {($\mathit{pres\_min}$)}. In addition, we have information on 
traffic flow {($\mathit{traffic}$)} averaged per day and per street, together with the maximum and minimum per day from 800 streets surrounding the measuring station.
Both traffic and pollution data are available \burl{https://datos.madrid.es/portal/site/egob} provided by the Madrid city council. 

{As noted before, and due to complex chemical reactions, NO2 is cross linked with O3 and other pollutants through non-linear forms in such a way that any increase or reduction in the level of O3 affects the level of NO2. In this line, we consider the latter as a response variable in our model to fully address these type of relationships for further control strategies.
}

Table \ref{tab:data} shows a description and summary statistics of the continuous covariates (before {standardisation} to $[0,1]$). The year from 2016--2019 is coded as 0/1 dummy variables with 2016 as a reference category.

\begin{table}[htbp] \renewcommand{\arraystretch}{1}
\centering\begin{tabular}{c|cccc}
  \hline\hline
Variable & Description & Mean & Std & Min/Max \\ 
  \hline
$\mathit{co}$ & carbon monoxide & 0.41 & 0.25 & 0.00/2.40 \\ 
  $\mathit{o3}$ & ozone & 39.57 & 20.29 & 0.00/89.00 \\ 
  $\mathit{prec}$ & precipitation & 1.01 & 3.35 & 0.00/28.50 \\ 
  $\mathit{temp}$ & average~temperature & 16.40 & 7.90 & 2.10/32.90 \\ 
  $\mathit{vel}$ & average ~wind speed & 1.80 & 1.00 & 0.00/6.40 \\ 
  $\mathit{racha}$ & wind gust speed & 8.94 & 3.28 & 1.90/26.10 \\ 
  $\mathit{pres\_max}$ & maximum~pressure & 943.20 & 5.58 & 922.00/967.30 \\ 
  $\mathit{pres\_min}$ & minimum~pressure & 938.75 & 6.17 & 915.00/957.40 \\ 
  $\mathit{traffic}$ & average~traffic flow & 776.22 & 147.36 & 249.63/1210.22 \\ 
   \hline\hline
\end{tabular}

\caption{\footnotesize{Description and summary statistics of continuous covariates (before standardisation to $[0,1]$ in the data set. The average traffic is measured as daily average from  approximately 800 segments of traffic flow around the Carmen station. In addition, we have the year from 2016--2019 coded as 0/1 dummy variables with 2016 as a reference category.}}\label{tab:data}
\end{table}

\begin{figure}[htbp]
\centering\includegraphics[width=0.48\textwidth,angle=0]{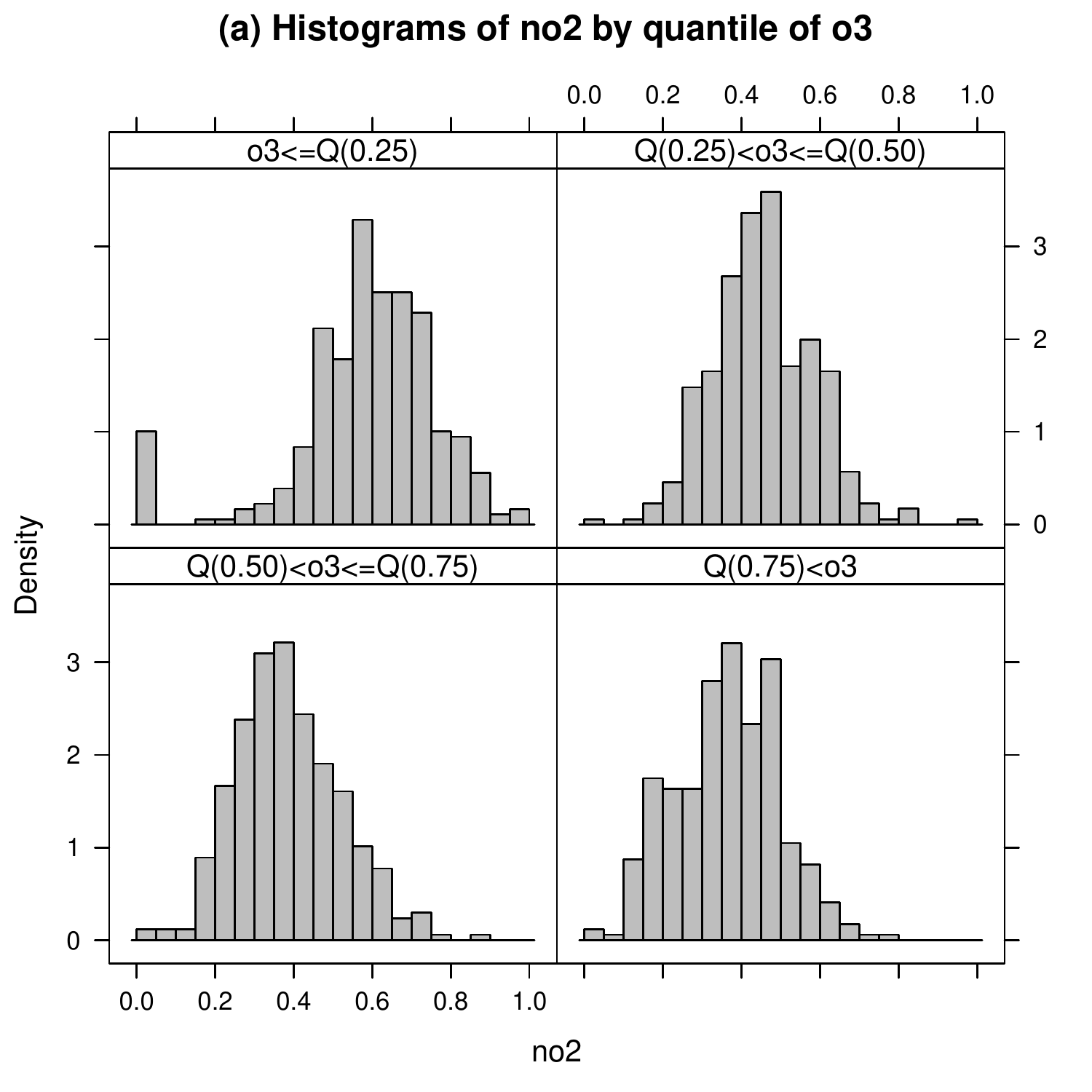}
\centering\includegraphics[width=0.48\textwidth,angle=0]{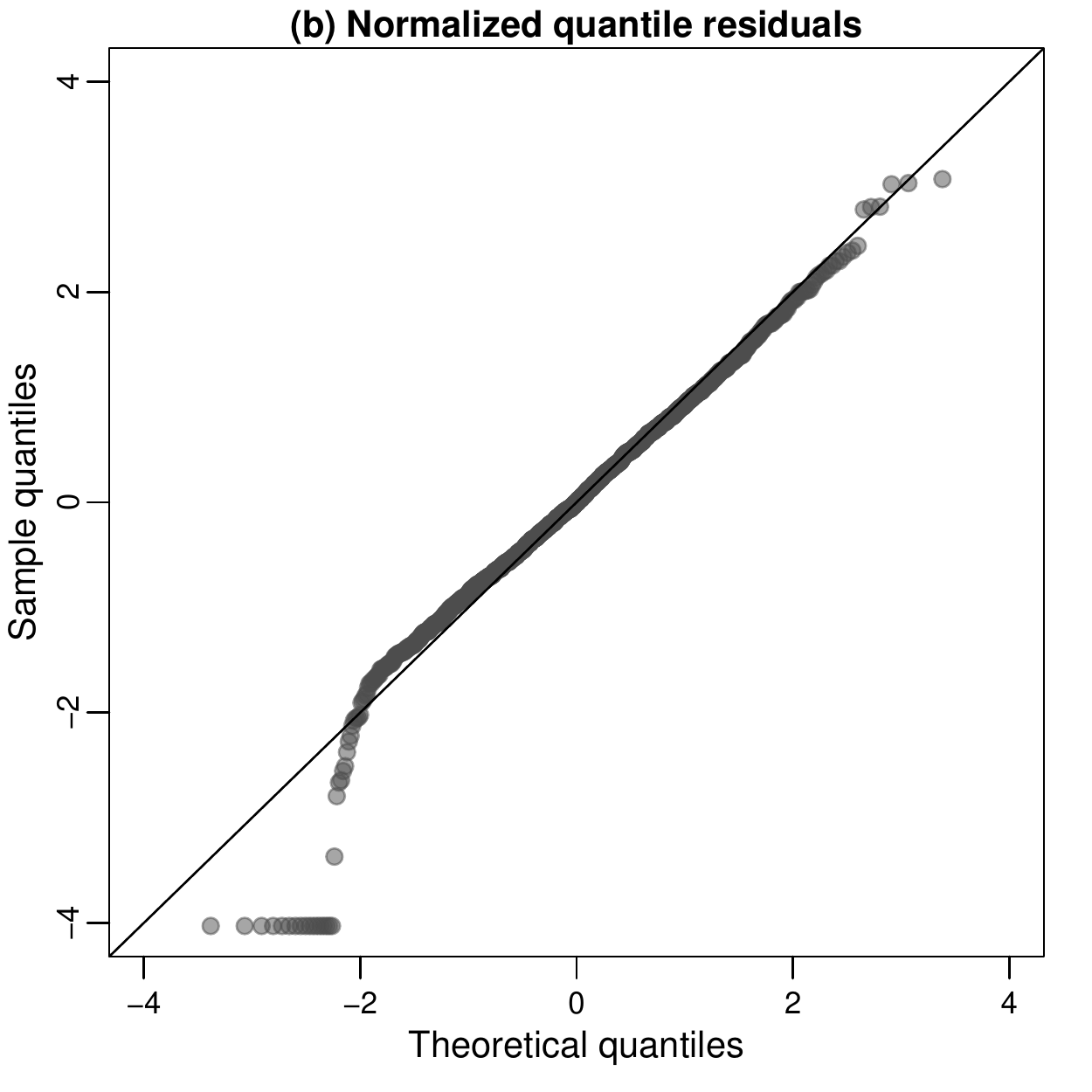}\\
\centering\includegraphics[width=0.48\textwidth,angle=0]{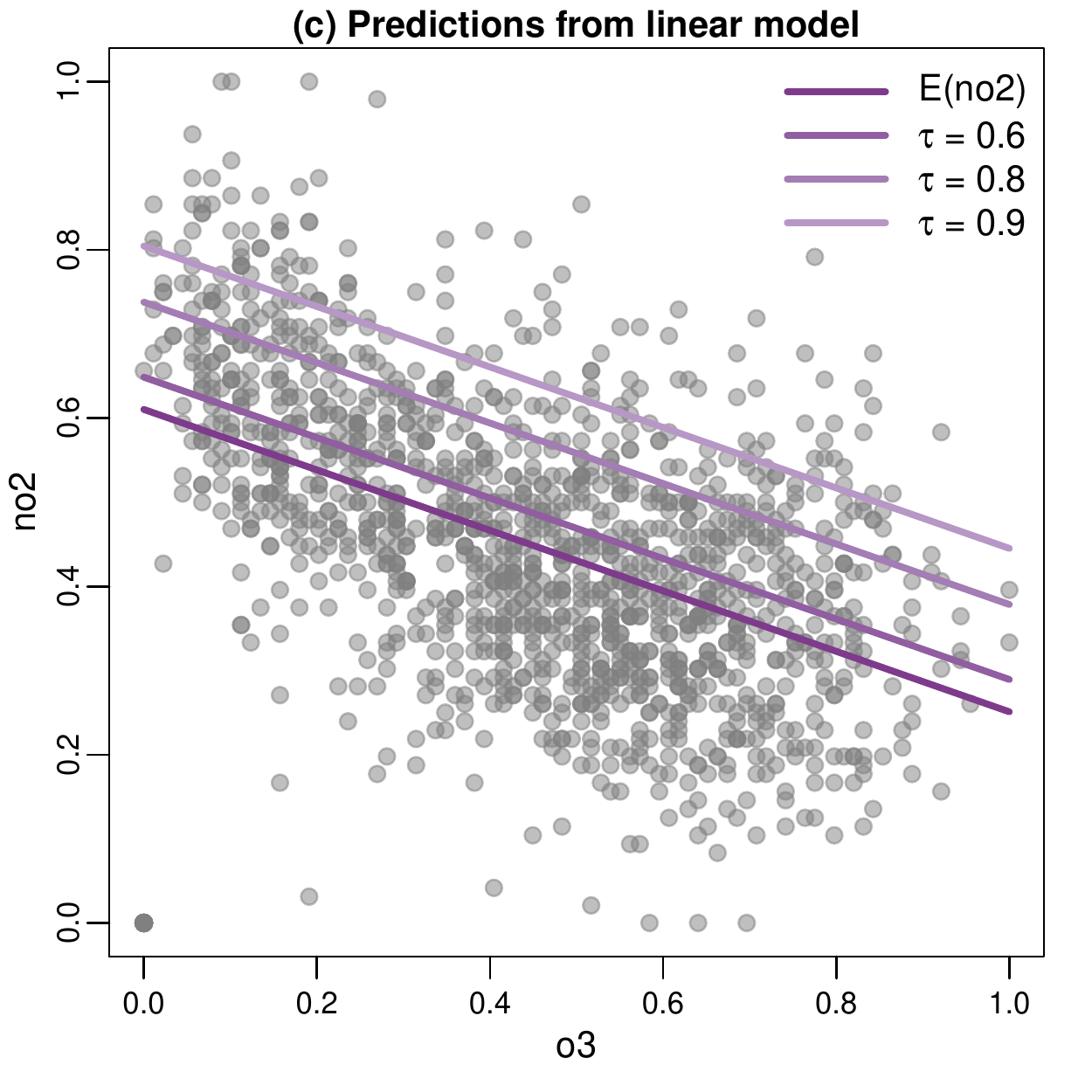}
\centering\includegraphics[width=0.48\textwidth,angle=0]{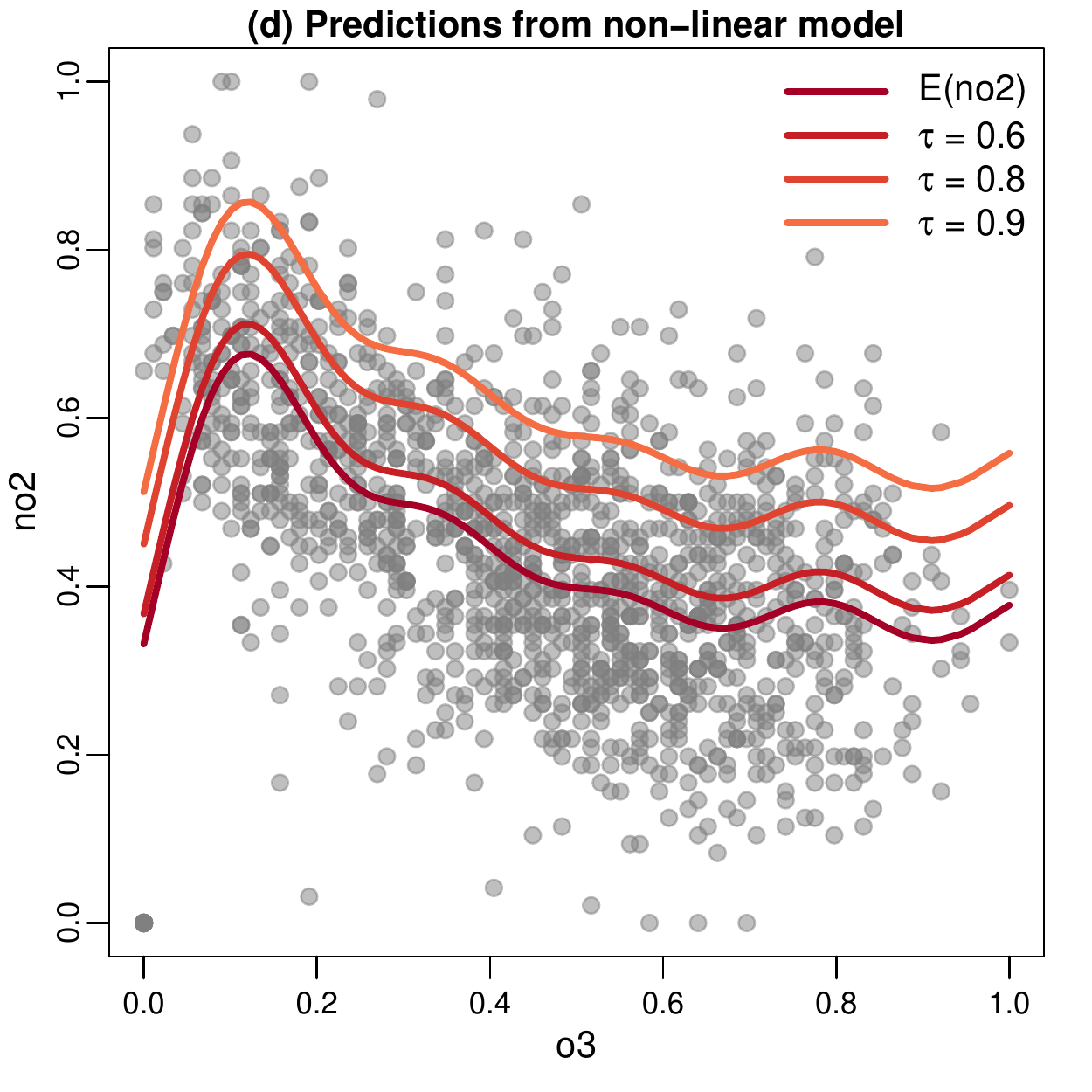}
\caption{\footnotesize{Results from univariate preliminary analyses with $\mathit{o3}$ as predictor. Panel (a) shows histograms of the response $\mathit{no2}$ according to the $\mathit{o3}$-quartiles, i.e.~where $Q(\tau)$, $\tau\in\lbrace 0.25,0.50,0.75\rbrace$ is the corresponding quartile of the empirical $\mathit{o3}$ distribution. Panel (b) shows normalised quantiles residuals from the univariate Gaussian regression model. Panels (c) and (d) show the predicted expectation $\dsE(\mathit{no2})$ as well as threshold quantiles $\tau\in\lbrace  0.6,0.8,0.9\rbrace$ obtained from a linear and non-linear Gaussian model, respectively.}}
\label{fig:prelim:no2:o3}
\end{figure}

\begin{figure}[htbp]
\centering\includegraphics[width=0.48\textwidth,angle=0]{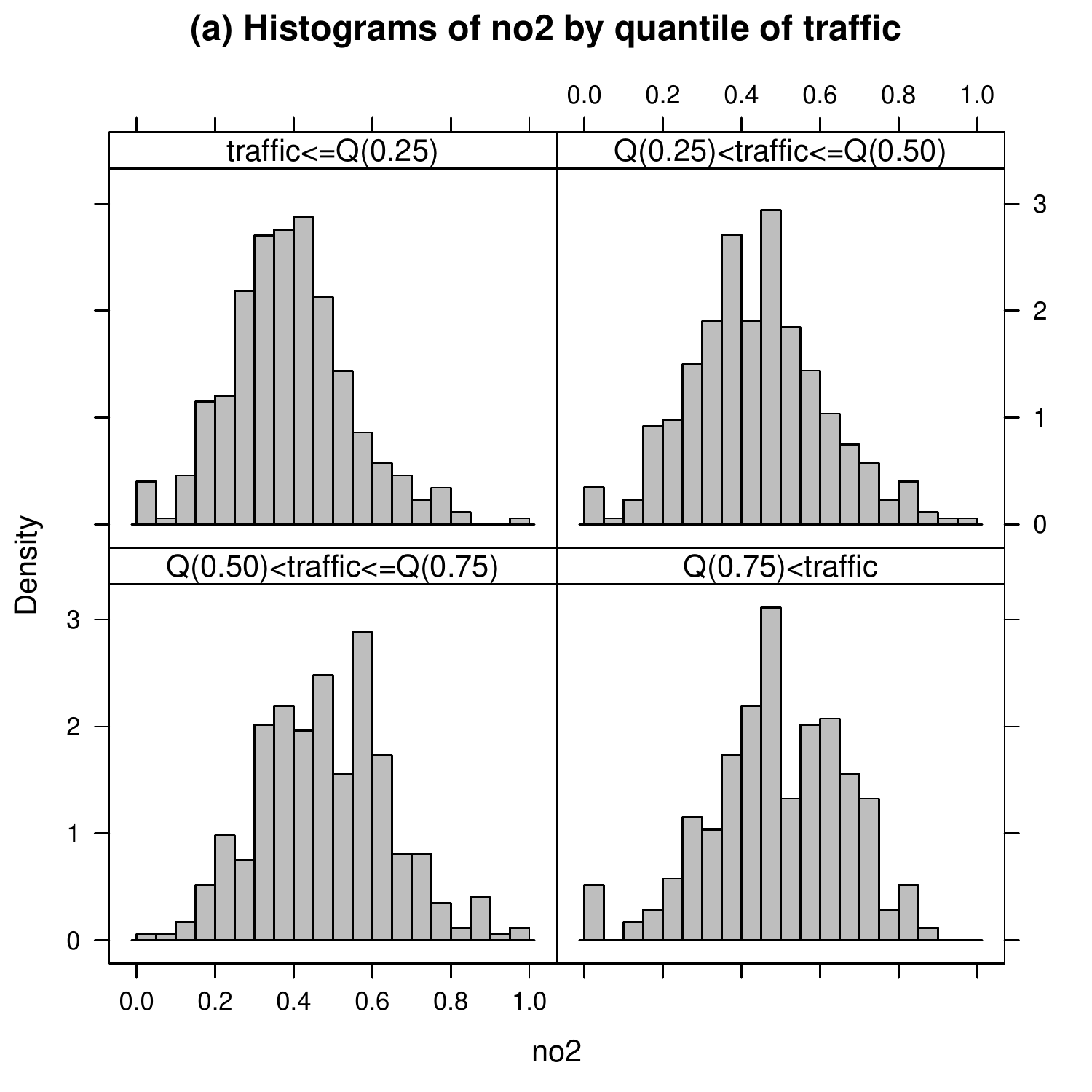}
\centering\includegraphics[width=0.48\textwidth,angle=0]{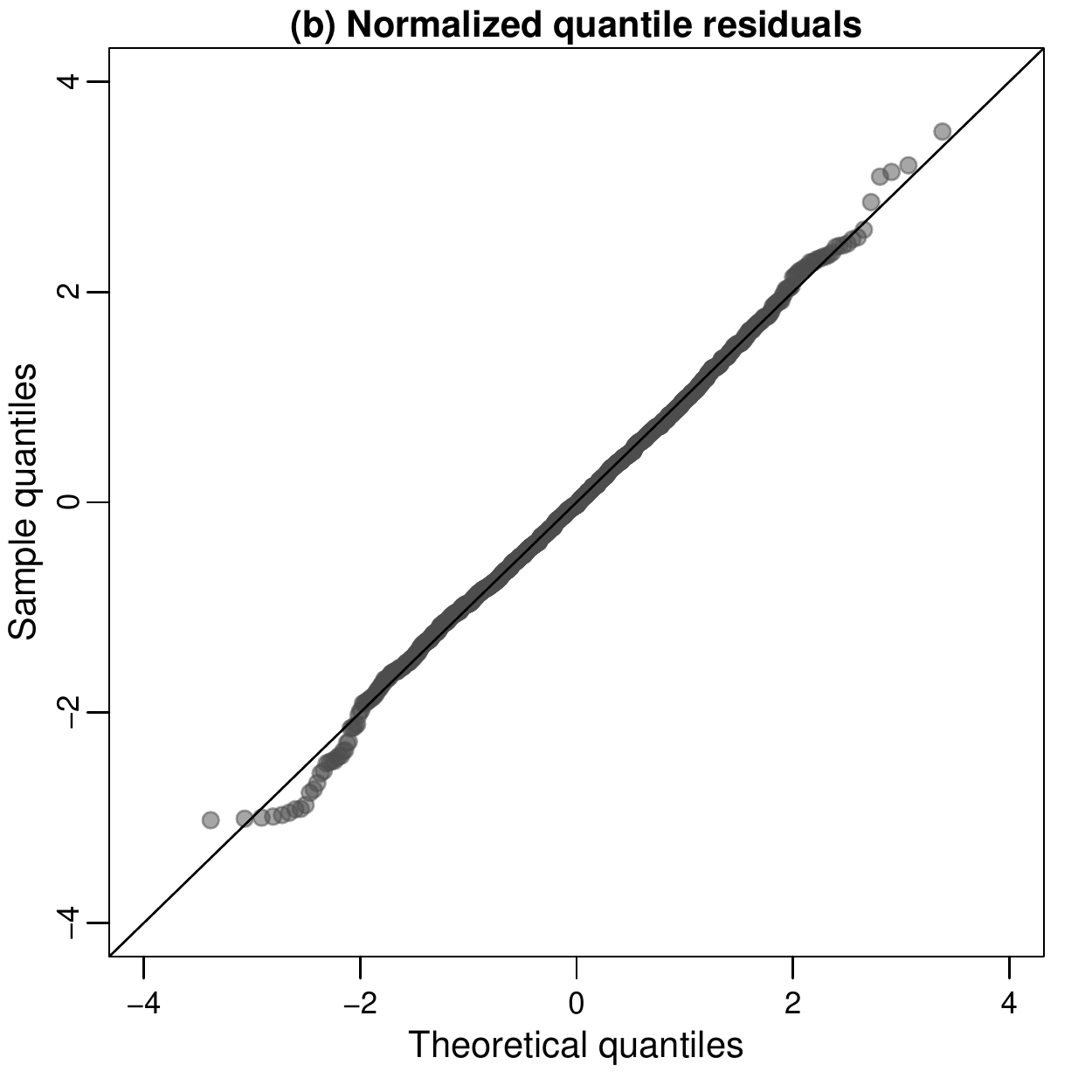}\\
\centering\includegraphics[width=0.48\textwidth,angle=0]{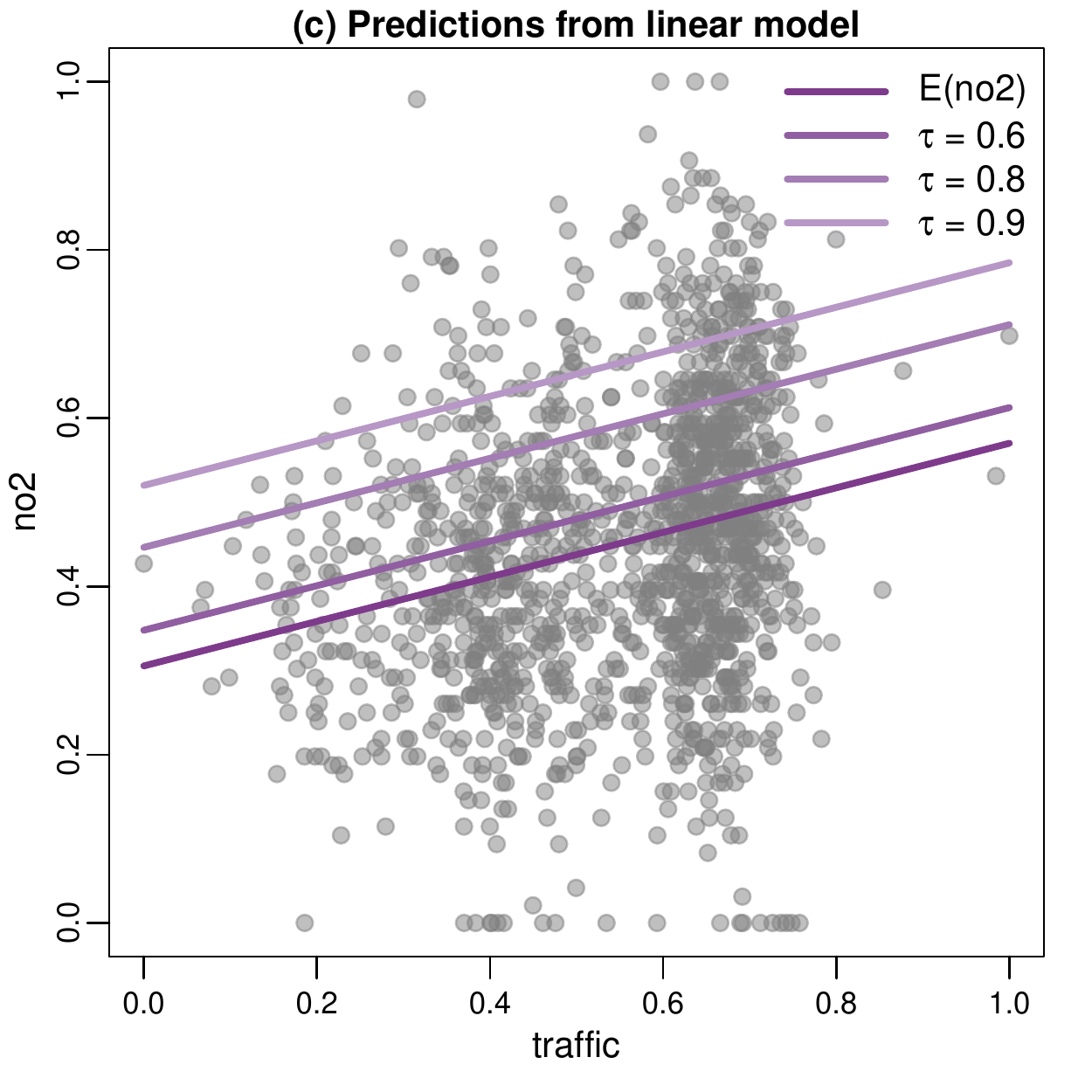}
\centering\includegraphics[width=0.48\textwidth,angle=0]{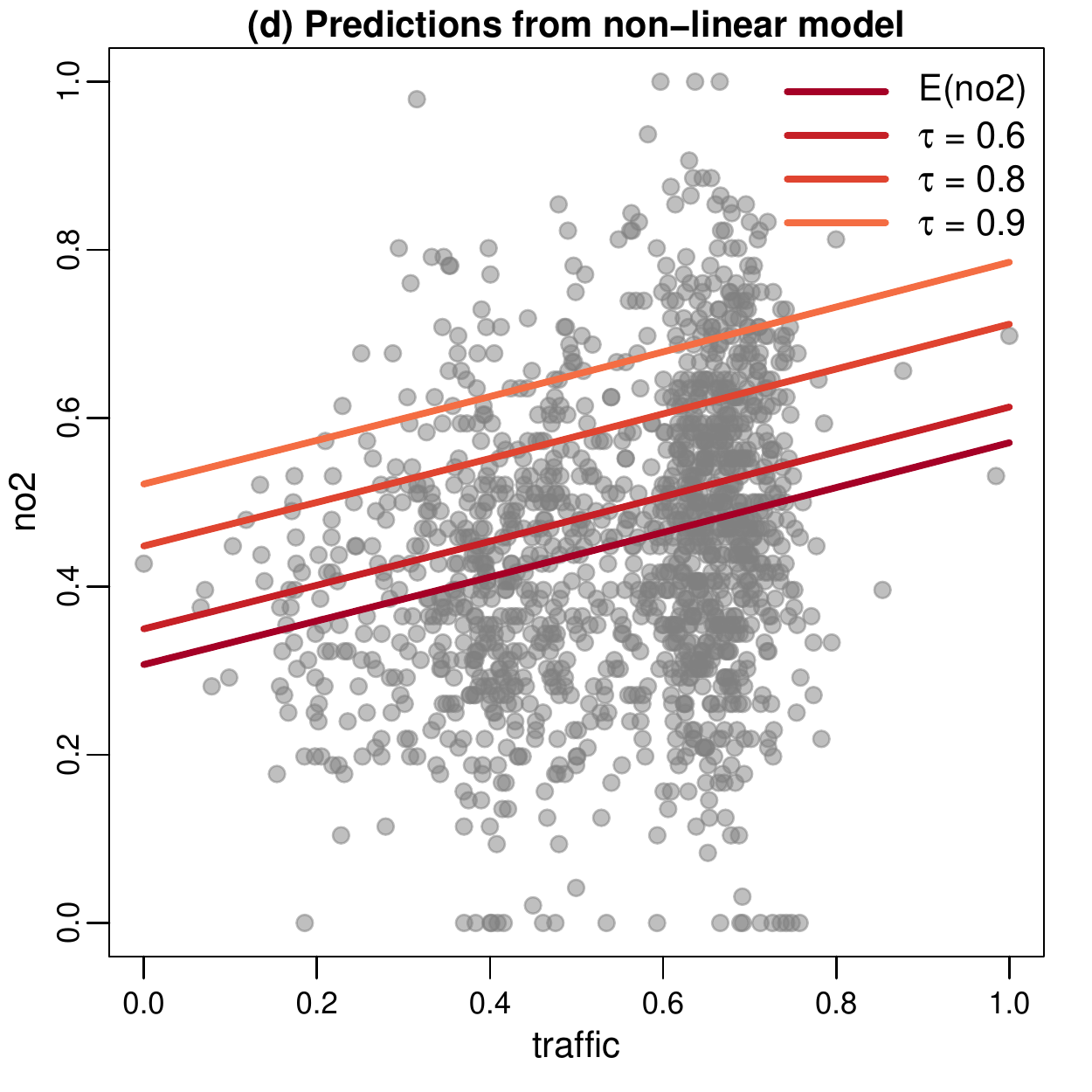}
\caption{\footnotesize{Results from univariate preliminary analyses with $\mathit{traffic}$ as predictor. Panel (a) shows histograms of the response $\mathit{no2}$ according to the $\mathit{traffic}$-quartiles, i.e.~where $Q(\tau)$, $\tau\in\lbrace 0.25,0.50,0.75\rbrace$ is the corresponding quartile of the empirical $\mathit{traffic}$ distribution. Panel (b) shows normalised quantiles residuals from the univariate Gaussian regression model. Panels (c) and (d) show the predicted expectation $\dsE(\mathit{no2})$ as well as threshold quantiles $\tau\in\lbrace  0.6,0.8,0.9\rbrace$ obtained from a linear and non-linear Gaussian model, respectively.}}
\label{fig:prelim:no2:traffic}
\end{figure}

Figure \ref{fig:prelim:no2:o3} depicts some preliminary univariate analyses with $\mathit{o3}$ as predictor. In particular, we show histograms of the response $\mathit{no2}$ according to the $\mathit{o3}$-quartiles, i.e.~where $Q(\tau)$, $\tau\in\lbrace 0.25,0.50,0.75\rbrace$ is the corresponding quartile of the empirical $\mathit{o3}$ distribution in panel (a). We also show normalised quantile residuals from the univariate Gaussian regression model (panel (b)). Then, we finally depict the predicted expectation $\dsE(\mathit{no2})$ as well as threshold quantiles $\tau\in\lbrace  0.6,0.8,0.9\rbrace$ obtained from a linear and non-linear Gaussian model, respectively (panels (c) and (d)).
In the same line, Figure \ref{fig:prelim:no2:traffic} shows some initial results when using $\mathit{traffic}$ as a predictor {in a univariate mean regression model}. Looking at {these two} figures, we note that  {depending on the predictor values, NO2 distributions differ (see panels (a)), {and looking at the tails we note that NO2 distributions are non-Gaussian} (see panels (b)).} In addition, the quantiles from linear and non-linear regression are parallel, which is not appropriate because conditional on traffic, the variance of NO2 is increasing with increasing traffic. Finally, we observe that the predictor O3 has a clear non-linear effect on NO2, while the one for traffic is rather linear.

\section{Bayesian effect selection for additive quantile regression}\label{sec:effselQR}

We first review Bayesian quantile regression using the auxiliary likelihood approach of~\citet{YuMoy2001} with latent Gaussian representation for computational feasibility, before we outline the normal beta prime spike and slab (NBPSS) prior for the quantile-specific regression coefficients. We also study effect decomposition into linear and respective non-linear effect parts as well as a feasible and interpretable way to eliciting hyperparameters of the NBPSS prior.

\subsection{Bayesian quantile regression with auxiliary likelihood}

 Let $(y_i, \nuvec_i)$, $i=1,\ldots,n$ denote $n$ independent observations on a continuous response variable $Y\in\mathcal{Y}\subset\mathds{R}$, and  $\nuvec$ the covariate vector comprising different types of covariate information such as discrete and continuous covariates or spatial information, see Section~\ref{subsec:semipred} for details. We then consider the model formulation 
\[
y_i=\eta_{i,\tau}+\varepsilon_{i,\tau}, \label{eq:M1}
\]
where $\eta_{i,\tau}$ is a structured additive predictor~\citep{FahKneLan2004} for a specific conditional quantile $\tau$ and $\varepsilon_{i,\tau}$ is an appropriate error term. Rather than assuming a zero mean for the errors as in classical mean regression, in quantile regression  we assume that the $\tau$th quantile of the error term is zero, i.e., $F_{\varepsilon_{i,\tau}}(0) =\tau$, where $F_{\varepsilon_{i,\tau}}(\cdot)$ denotes the cumulative distribution function of the $i$th error term. This assumption implies that the predictor $\eta_{i,\tau}$ specifies the $\tau$th quantile of $y_i$ and, as a consequence, the regression effects can be directly interpreted on the quantiles of the response distribution. Although we only consider here three specific quantiles of interest, estimation for a dense set of quantiles would allow us to {characterise} the complete distribution of the responses in terms of covariates~\cite[see, e.g.][]{BonReiWan2010,SchEil2013,RodDorFan2019}.

In the Bayesian framework, a specific distribution for the errors is required to facilitate full posterior inference. For this purpose, the asymmetric Laplace distribution (ALD) $y_i\sim\mbox{ALD}(\eta_{i,\tau},\delta^{2},\tau)$ with location predictor $\eta_{i,\tau}$, scale parameter $\delta^{2}$, asymmetry parameter $\tau$ and density
\[
p(y_i|\eta_{i,\tau},\delta^{2},\tau)=\tau(1-\tau)\delta^{2}\exp(-\delta^{2}\rho_\tau(y_i-\eta_{i,\tau}))
\]
is particularly useful~\citep{YuMoy2001}, since it can be shown to yield posterior mode estimates that are equivalent to the minimisers of \eqref{eq:awad}.
However,  estimation with the ALD directly is not straightforward due to the non-differentiability of the check function $\rho_\tau(\cdot)$ at zero. Instead, to make Bayesian inference efficient, several authors considered re-writing the ALD as a scale mixture of two Gaussian distributions~\citep{ReeYu2009,KozKob2011,YueRue2011,LumGel2012,WalKneLanYue2012}. Specifically, \citet{Tsi2003} show that 
\[\label{eq:M1a}
Y=\eta+\xi W + \sigma Z\sqrt{\delta^{-2} W},\quad W\sim\ExpD(\delta^{2}),\;Z\sim\ND(0,1)
\]
has an ALD distribution, where $\xi=\tfrac{1-2\tau}{\tau(1-\tau)}$, $\sigma=\tfrac{2}{\tau(1-\tau)}$ are two scalars depending on the quantile level $\tau$; and the random variables $W$, $Z$ are independently distributed according to exponential ($\ExpD(\delta^{2})$ with rate $\delta^2$) and standard Gaussian ($\ND(0,1)$) distributions, respectively. We assume a conjugate gamma prior distribution $\delta^2\sim\GaD(a_\delta,b_\delta)$. This representation as a mixture facilitates an easy way to construct Gibbs sampling for MCMC inference, see Section~\ref{subsec:post} below.

{We last note} that other than the ALD are used in the literature to perform Bayesian quantile regression. For instance, \citet{KotKrn2009} construct a generic class of
semi-parametric and non-parametric distributions for the likelihood using Dirichlet process mixture models, while \citet{ReiBonWan2010} consider a flexible infinite mixture of Gaussian combined with a stick-breaking construction for the priors.
However, we follow \citet{YueRue2011} along with the ALD since it 
 facilitates Bayesian inference and
computation in the  additive models of this paper; and because \citet{YueRue2011} show empirically, that the ALD is flexible enough to capture various deviations from normality (such as skewness, heavy tails, etc.).

\subsection{Semi-parametric predictors}\label{subsec:semipred}

In the following we present the formulation in the most general case to have a clear image of the generality of our proposal. Later, when we restrict to the application we will set a particular case, for example, all our covariates (except year-specific dummies) are subject to selection and there are no one free of selection. Also, we only consider splines for continuous covariates since we do not have random effects/spatial effects.

The predictors $\eta_{i,\tau}$ are decomposed into $\eta_{i,\tau}=\eta_{i,\tau}^{\mathrm{in}} +\eta_{i,\tau}^{\mathrm{sel}}$, i.e. a sub-predictor $\eta_{i,\tau}^{\mathrm{sel}}$ being subject to explicit effect selection via spike and slab priors, and a second sub-predictor  $\eta_{i,\tau}^{\mathrm{in}}$ containing effects not subject to selection. {We assume} that $\eta_{i,\tau}^{\mathrm{sel}}$ and $\eta_{i,\tau}^{\mathrm{in}}$ are disjoint. The separation into two subsets of effects allows us to include specific covariate effects mandatory in the model (e.g.~based on prior knowledge or since these represent confounding effects that have to be included in the model). We then model the predictors in a structured additive fashion along the lines of~\citet{FahKneLan2004}
\[\label{eq:M2}
 \eta_{i,\tau} =   \sum_{l=1}^{L_\tau}f_{l,\tau}^{\mathrm{in}}(\nuvec_i) + \sum_{j=1}^{J_\tau}f_{j,\tau}^{\mathrm{sel}}(\nuvec_i),
\]
where the effects $f_{j,\tau}^{\mathrm{sel}}(\nuvec_i)$ and $f_{l,\tau}^{\mathrm{in}}(\nuvec_i)$ represent various types of flexible functions depending on (different subsets of) the covariate vector $\nuvec_i$ that are to be selected via spike and slab priors and those not subject to selection, respectively.  In the following, we focus on the specification of effect selection priors for $f_{j,\tau}^{\mathrm{sel}}(\nuvec_i)$ since in our application no specific functional forms of one of the nine covariates should a priori be excluded from selection.  For a short period of time, it is assumed that there is no much uncertainty amongst the years, and we can savely consider the $\mathit{year}$ as a fixed effect covariate. Estimation of the corresponding coefficients is handled as done in~\cite{WalKneLanYue2012}.

In STAQ models it is  assumed that each effect $j$ in quantile $\tau$, $f_{j,\tau}$,  can be modelled as
\[
 f_{j,\tau}(\nuvec_i) = \sum_{d=1}^{D}\beta_{j,k,d}B_{j,\tau,d}(\nuvec_i),
\]
where $B_{j,\tau,d}(\nuvec_i)$, $d=1,\ldots,D$ are appropriate basis functions and $\betavec_{j,\tau}=(\beta_{j,\tau,1},\ldots,\beta_{j,\tau,D})'$ is the vector of unknown basis coefficients. 

\vspace{0.5em}
\paragraph{Scalar parameter expansion for effect selection}
To decide now for the overall relevance of $f_{j,\tau}$ in~\eqref{eq:M1}, we follow~\citet{KleCarKneLanWag2021} and others and {reparameterise} the equation above to
\[\label{eq:genericeffect}
 f_{j,\tau}(\nuvec_i) = \zeta_{j,\tau}\sum_{d=1}^{D}\tilde\beta_{j,\tau,d}B_{j,\tau,d}(\nuvec_i),
\]
where now $\betatildevec_{j,\tau}=(\tilde\beta_{j,\tau,1},\ldots,\tilde\beta_{j,\tau,D})'$ is the {  (standardised) vector} of basis coefficients, and $\zeta_{j,\tau}$ is a scalar importance parameter. The latter is assigned a spike and slab prior in the next subsection (more precisely we place the prior on the squared importance parameter $\zeta_{j,\tau}^2$). This allows us to remove the effect from the predictor for $\zeta_{j,\tau}$ close to zero, while the effect is considered to be of high importance if $\zeta_{j,\tau}$ is large in absolute terms. Hence, instead of doing selection directly on the (possibly high-dimensional) vector $\betavec_{j,\tau}$ we can boil done the problem of selection on scalar parameters.  This is reasonable due to the aim to select an effect with a corresponding vector $\betavec_{j,\tau}$ as a whole rather than single coefficients.

\vspace{0.5em}
\paragraph{Relevant examples}
{Due to the linear basis representation, the vector of function evaluations $\fvec_{j,\tau}=(f(\nuvec_{j,\tau,1}),$
$\ldots,f(\nuvec_{j,\tau,n}))'$ can now be written} as $\fvec_{j,\tau}=\zeta_{j,\tau}\mB_{j,\tau}\betatildevec_{j,\tau}$ where $\mB_{j,\tau}$ is the ($n\times D$) design matrix arising from the evaluation of the basis functions $B_{j,\tau,d}(\nuvec_i)$, $d=1,\ldots,D$ at the observed  $\nuvec_1,\ldots,\nuvec_n$.
While the STAR/STAQ framework enables a variety of different effect types, in the following we briefly discuss some details on linear and non-linear effects of univariate continuous covariates only (as these are the ones important in our application), while we refer the reader to~\cite{Woo2017} for more terms, such as spatial effects or random effects.
{The basis functions $\mB_{j,\tau}$ depend very much on the type of effect (linear/non-linear) we are considering for the covariates.} 

For \emph{linear effects} of continuous covariates, the columns of the design matrix $\mB_{j,\tau}$ are  equal to the different covariates. For binary/categorical covariates, the basis functions represent the chosen coding, e.g.~dummy or effect coding and the design matrix then consists of the resulting dummy or effect coding columns. 

{For} a \emph{non-linear effect of a continuous covariate} we employ Bayesian P-splines~\citep{LanBre2004}. The $i$th row of the design matrix $\mB_{j,\tau}$ then contains the B-spline basis functions $B_{j,k,1}(x_i),\ldots,B_{j,k,D}(x_i)$ evaluated at the observed covariate value $x_i$. If not stated otherwise, we will use cubic B-splines with seven inner knots (resulting in effects of dimension $D=9$). This choice turns out to be sufficiently large in our case. We compared this to $D=22$ (20 inner knots) and $D=42$ (40 inner knots) following the default values considered in~\citet{LanBre2004}  and~\citet{EilMar1996} but found the smaller number to still ensure enough flexibility.

\subsection{Hierarchical spike and slab prior for effect selection}\label{subsec:NBPSS}

\vspace{0.5em}
\paragraph{Constraint prior for regression coefficients}

To enforce specific properties such as smoothness or shrinkage, we assume multivariate Gaussian priors for the scaled basis coefficients. {Thus we consider the following  prior for the vector} of standardised basis coefficients $\tilde\betavec_{j,\tau}$ 
\[\label{eq:M3}	
 p(\betatildevec_{j,\tau})\propto\exp\left(-\frac{1}{2}\betatildevec_{j,\tau}'\mK_{j,\tau}\betatildevec_{j,\tau}\right)\mathds{1}\left\lbrack\mA_{j,\tau}\betatildevec_{j,\tau}=\nullvec\right\rbrack,
\]
where $\mK_{j,\tau}\in\dsR^{D_{j,\tau}\times D_{j,\tau}}$ denotes the prior precision matrix implementing the desired smoothness properties, 
and the indicator function $\mathds{1}[\mA_{j,\tau}\betavec_{j,\tau}= \nullvec]$ is included to enforce linear constraints on the regression coefficients via the constraint matrix $\mA_{j,\tau}$. The latter is typically used to remove identifiability problems from the additive predictor (e.g.~by centering the additive components of the predictor) but can also be used to remove the partial impropriety from the prior that comes from a potential rank deficiency of  $\mK_{j,\tau}$ with $\rank(\mK_{j,\tau})=\kappa_{j,\tau}\le D_{j,\tau}$.
 Here, we specifically assume that the constraint matrix $\mA_{j,\tau}$ is chosen such that all rank deficiencies in $\mK_{j,\tau}$ are effectively removed by setting
$
 \mA_{j,\tau}=\spa\left(\ker(\mK_{j,\tau})\right),
$ 
where $\ker(\mK_{j,\tau})$ denotes the null space of $\mK_{j,\tau}$ and $\spa\left(\ker(\mK_{j,\tau})\right)$ is a representation of the corresponding basis. 

{To select the prior precisions $\mK_{j,\tau}$, we again consider the type of effect for the covariates.} For \emph{linear effects}, we choose $\mK_{j,\tau}=\mI$, while for a \emph{non-linear effect of a continuous covariate} we employ a second order random walk prior in all our empirical applications. Removing all rank deficiencies does not only remove the non-propriety from the prior, but also allows to make the relation between the original and the parameter expansion more explicit and to perform effect decomposition for the components of the additive predictor.

\vspace{0.5em}
\paragraph{Effect decomposition}
{With the above assumption}, for Bayesian P-splines with a second order random walk prior, the rank of the prior precision matrix is $\kappa_{j,\tau}=D-2$ and the null space corresponds to constant and linear effects. Applying the constrained prior allows to select  linear effects and non-linear deviations separately. In general, an effect $f_{j,\tau}(\nuvec)$ can be decomposed into one {unpenalised} component $f_{j,k,\unpen}(\nuvec)$ that corresponds to the null space of the prior precision matrix and the penalised complement $f_{j,k,\pen}(\nuvec)$ 
\[
 f_{j,\tau}(\nuvec)=f_{j,k,\unpen}(\nuvec)+f_{j,k,\pen}(\nuvec).
\]
To achieve separate effect selection for the two components of $f$, we assign distinct spike and slab priors~\citep{KleCarKneLanWag2021,rossell2019}. 

\vspace{0.5em}
\paragraph{Scaled beta prime spike and slab prior on squared importance parameter}
We follow \cite{KleCarKneLanWag2021} and place the following hierarchical prior specification on the squared importance parameter $\zeta_{j,\tau}^2$:
\begin{equation}
\label{eq:M5}	
 \begin{aligned}
 \zeta_{j,\tau}^2|\gamma_{j,\tau},\psi_{j,\tau}^2 & \sim\GaD\left(\frac{1}{2},\frac{1}{2r_{j,\tau}(\gamma_{j,\tau}) \psi_{j,\tau}^2}\right) \\
 \gamma_{j,\tau}|\omega_{j,\tau}   & \sim \BerD(\omega_{j,\tau})\\
 \psi_{j,\tau}^2    & \sim \IGD(a_{j,\tau},b_{j,\tau})\\
 \omega_{j,\tau}    & \sim \BetaD(a_{0,j,\tau},b_{0,j,\tau})\\
 r_{j,\tau}\equiv r(\gamma_{j,\tau}) &=\begin{cases} r_{j,\tau}>0 \mbox{ small } & \gamma_{j,\tau}=0\\
                          1  & \gamma_{j,\tau}=1,
            \end{cases}
 \end{aligned}
\end{equation}
where $\GaD(a,b)$ denotes a gamma distribution with shape and scale parameters $a,b$, $\BerD(p)$ is a Bernoulli distribution with success probability $p$, $\IGD(a,b)$ is an inverse gamma distribution with parameters $a,b$ and $\BetaD(a,b)$ reads a beta distribution with shape parameters $a,b$.

The general idea of this prior hence relies on a mixture of one prior concentrated around zero such that it can effectively be thought of as representing zero (the spike component) and a more dispersed, mostly noninformative prior (the slab), and specified via the hierarchy. Specifically, the scale parameter $\psi_{j,\tau}^2$ determines the prior expectation of $\zeta_{j,\tau}^2$, which is $\psi_{j,\tau}^2$ for $\gamma_{j,\tau}=1$ and $r_{j,\tau} \psi_{j,\tau}^2$ for $\gamma_{j,\tau}=0$ with $r_{j,\tau}\ll1$ being a fixed small {value; hence} the indicator $\gamma_{j,\tau}$ determines whether a specific effect $\betavec_{j,\tau}=\tau_{j,\tau} \betatildevec_{j,\tau}$ is included in the model ($\gamma_{j,\tau}=1$) or excluded from the model ($\gamma_{j,\tau}=0$). The parameter $\omega_{j,\tau}$ is the prior probability for an effect being included in the model and the remaining parameters $a_{j,\tau}$, $b_{j,\tau}$, $a_{0,j,\tau}$, $b_{0,j,\tau}$ and $r_{j,\tau}$ are hyperparameters of the spike and slab prior. Prior elicitation for these parameters {comes in detail} in Section~\ref{subsec:PriEli}, while the other choices follow those of~\cite{KleCarKneLanWag2021} and are derived from the theoretical results investigated in this paper and~\citet{PerPerRam2017}.

\subsection{Prior elicitation}\label{subsec:PriEli}
We now briefly outline prior elicitation for the hyperparameters $a_{j,\tau}$, $b_{j,\tau}$, $_{0,j,\tau}$, $b_{j,0,\tau}$ and $r_{j,\tau}$. 
\citet{PerPerRam2017} show that the moments of order less than $a_{j,\tau}$ exist and the variance decreases with $a_{j,\tau}$. Furthermore, for small values of $a_{j,\tau}$, the spike and the slab component will overlap such that moves from $\gamma_{j,\tau}=0$ to $\gamma_{j,\tau}=1$ are possible. We follow previous authors and
set $a_{j,\tau}=5$ as a default.

Since we do not have specific prior information about the prior inclusion probability of effects, we use $a_{j,0,\tau}=b_{0,j,\tau}=1$, which corresponds to a flat prior on the unit interval. In general we note that prior assumptions can be implemented, since $\dsP(\gamma_{j,\tau}=1|a_{0,j,\tau},b_{0,j,\tau})=a_0/(a_{0,j,\tau}+b_{0,j,\tau})$. 

For the elicitation of $b_{j,\tau}$ and $r_{j,\tau}$, we use a more efficient procedure of what is done in~\citet{KleCarKneLanWag2021}, an approach which itself is based on prior work of~\citet{SimRueMarRieSor2017} and \citet{KleKne2016}. More precisely, we consider marginal probability statements on the supremum norm of $f_{j,\tau}$,  $\supnorm{f_{j,\tau}}$,  and conditional on the status of the inclusion/exclusion parameter $\gamma_{j,\tau}$. Given $\gamma_{j,\tau}=1$ (inclusion of the effect), the marginal distribution of $f_{j,\tau}(\nuvec)$ does no longer depend on $r_{j,\tau}$, such that the parameter $b_{j,\tau}$ can be determined from
\begin{equation}\label{eq:cond:hyper2}
 \dsP\left(\left.\supnorm{f_{j,\tau}}\,\leq c_{j,\tau}\,\right|\,\gamma_{j,\tau}=1\right) = \alpha_{j,\tau}.
\end{equation}
This is the probability that the supremum norm of an effect is smaller than a pre-specified level $c_{j,\tau}$. 
 Given $\gamma_{j,\tau}=1$,  $\alpha_{j,\tau}$ and $c_{j,\tau}$ should be small, reflecting the prior beliefs of the unlikely event that $\supnorm{f_{j,\tau}}$  was smaller than $c_{j,\tau}$ if it was indeed an informative effect to be included into the predictor. Both the level $c_{j,\tau}$ and the prior probability $\alpha_{j,\tau}$ have to be specified by the analyst according to concrete prior beliefs. To derive $r_{j,\tau}$, we proceed similarly but {considering} the reverse scenario summarised via the probability
\begin{equation}\label{eq:cond:hyper1}
 \dsP\left(\left.\supnorm{f_{j,\tau}}\,\leq c_{j,\tau}\,\right|\,\gamma_{j,\tau}=0\right) = 1-\alpha_{j,\tau}
\end{equation}
now conditioning on non-inclusion. Since in this case the probability of not exceeding the threshold $c_{j,\tau}$ should be large, the probability is reversed to $1-\alpha_{j,\tau}$ for $\alpha_{j,\tau}$ still small. Note that the absolute value of the effects can be taken without loss of generality due to the centring constraint of each function to ensure identifiability.

The basic idea of \eqref{eq:cond:hyper2}--\eqref{eq:cond:hyper1} is that such prior statements can be much more easily elicited in applications. 
To now access these probabilities , we generate a large sample from the marginal distribution of $\sup_{\nuvec\in\calD} |f_{j,\tau}(\nuvec)|$ (the latter can be approximated via simulation from the full hierarchical model). While~\cite{KleCarKneLanWag2021} suggest solving \eqref{eq:cond:hyper2} and  \eqref{eq:cond:hyper1} with respect to $r_{j,\tau}$ and $b_{j,\tau}$ (which can be done independently) numerically via a time consuming optimization procedure, we note that this can be made much more efficient by making use of the scaling property of the scaled beta prime distribution:  As shown in~\citet{PerPerRam2017}, the marginal priors of spike and the slab components (i.e.~with $\psi_{j,\tau}^2$ integrated out) $p(\zeta_{j,\tau}^2|\gamma_{j,\tau})$ are scaled beta prime distributions with shape parameters 1/2 and $a_{j,\tau}$ and scale parameter
$2r_{j,\tau}b_{j,\tau}$. Hence, $2r_{j,\tau}b_{j,\tau}\zeta_{j,\tau}^2|\gamma_{j,\tau}$ has a beta prime distribution with shape parameters 1/2 and $a_{j,\tau}$ ($\mathcal{BP}(1/2,a_{j,\tau}$)) and computing the probability in \eqref{eq:cond:hyper2} is equivalent to computing  $$\dsP\left(\left.\supnorm{f_{j,\tau}}\,\leq c_{j,\tau}\,\right|\,\gamma_{j,\tau}=1\right) =\dsP\left(\left.\supnorm{\tilde\zeta_{j,\tau}\bvec'\tilde\betavec_{j,\tau}}\,\leq \frac{c_{j,\tau}}{ \sqrt{2 r_{j,\tau}b_{j,\tau}}}\,\right|\,\gamma_{j,\tau}=1\right),$$ where we have defined $f_{j,\tau}=\zeta_{j,\tau}\bvec'\tilde\betavec_{j,\tau}=\sqrt{2 r_{j,\tau}b_{j,\tau}}\tilde\zeta_{j,\tau}\bvec'\tilde\betavec_{j,\tau}$, and $\tilde\zeta_{j,\tau}$ is the square root of a $\mathcal{BP}(1/2,a_{j,\tau})$-distributed random variable. As a {consequence, no numerical optimization} is needed but it suffices to i) generate a large sample from $\supnorm{\tilde\zeta_{j,\tau}\bvec'\tilde\betavec_{j,\tau}}$, ii) determine its $\alpha_{j,\tau}$-quantile $q^\ast$ , and iii) set $r_{j,\tau}^\ast$ such that $c_{j,\tau}/r_{j,\tau}^\ast=q^\ast$. The equivalent procedure applies to solving \eqref{eq:cond:hyper1}. 

The parameters $\alpha_{j,\tau}$ and $c_{j,\tau}$ are easy to interpret and better accessible to the analyst as compared to all hyperparameters in~\eqref{eq:M5}. Their specific choice  can help to balance
between the true positive and false negative rates of effect selection. For instance, choosing
$\alpha_{j,\tau}$ and $c_{j,\tau}$ smaller, will yield more conservative, i.e. sparser models. Based on the simulation results of~\citet{KleCarKneLanWag2021} the default value of $\alpha_{j,\tau}=c_{j,\tau}=0.1$ is used in our empirical analysis.
Finally, all covariates subject to selection are standardised to the interval $[0,1]$ since this facilitates prior elicitation and comparability across the covariates.

\subsection{Posterior inference}\label{subsec:post}

One appealing property in our effect selection STAQ model compared to~\cite{KleCarKneLanWag2021} is the more efficient MCMC  sampling scheme that does not require any Metropolis-Hastings steps for the regression coefficients $\betavec_{j,\tau}$ but all steps can be {realised} in Gibbs updates as follows. Let $$\thetavec=\lbrace \betavec_{1,\tau},\ldots,\betavec_{J,\tau},\zeta_{1,\tau}^2,\ldots,\zeta_{J,\tau}^2,\gamma_{1,\tau},\ldots,\gamma_{J,\tau},\psi_{1,\tau},\ldots,\psi_{J,\tau},\omega_{1,\tau},\ldots,\omega_{J,\tau},w_1,\ldots,w_n,\delta^2\rbrace$$ be the set of all model parameters, then the MCMC sampler can be {summarised as follows}.

\vspace{0.5em}
\paragraph{MCMC Sampler for effect selection in STAQ models}
\textcolor{white}{xx}\\
\noindent At each MCMC sweep:\\
\noindent \underline{Step~1.} For $j=1,\ldots,J$ generate from $p(\betavec_{j,\tau}\,|\,\thetavec\backslash\betavec_{j,\tau})$.\\
\noindent \underline{Step~2.} For $j=1,\ldots,J$ generate from $p(\zeta_{j,\tau}\,|\,\thetavec\backslash\zeta_{j,\tau})$.\\
\noindent \underline{Step~3.} For $j=1,\ldots,J$ generate from $p(\gamma_{j,\tau}\,|\,\thetavec\backslash\gamma_{j,\tau})$.\\
\noindent \underline{Step~4.} For $j=1,\ldots,J$ generate from $p(\psi_{j,\tau}\,|\,\thetavec\backslash\psi_{j,\tau})$.\\
\noindent \underline{Step~5.} For $j=1,\ldots,J$ generate from $p(\omega_{j,\tau}\,|\,\thetavec\backslash\omega_{j,\tau})$.\\
\noindent \underline{Step~6.} For $i=1,\ldots,n$ generate from $p(w_i\,|\,\thetavec\backslash w_i)$. \\
\noindent \underline{Step~7.} Generate from $p(\delta^2\,|\,\thetavec\backslash\delta^2)$.

\vspace{0.9em}

The full conditional distributions for $\betavec_{j,\tau}$ are Gaussian distributions $\ND(\muvec_{j,\tau},\mSigma_{j,\tau})$ due to the conjugate model hierarchy implied by the location-scale mixture representation and the Gaussian priors for $\betavec_{j,\tau}$. The mean $\muvec_{j,\tau}$ and covariance $\mSigma_{j,\tau}$ are:
\begin{equation*}\begin{aligned}
 \muvec_{j,\tau}&=\mSigma_{j,\tau}^{-1}\left(\tfrac{\delta^2}{\sigma^2}\mB_{j,\tau}'\mD_w^{-1}(\yvec-\xi\wvec-(\etavec-\mB_{j,\tau}\betavec_{j,\tau}))\right)
\\
\mSigma_{j,\tau}&=\left(\mK_{j,\tau}+\tfrac{\delta^2}{\sigma^2}\mB_{j,\tau}'\mD_w^{-1}\mB_{j,\tau}\right)^{-1},
\end{aligned}\end{equation*}
where $\wvec=(w_1,\ldots,w_n)'$ and $\mD_w=\mbox{diag}(w_1,\ldots,w_n)$. 
At Step 2, we note that $p(\zeta_{j,\tau}^2|\betavec_{j,k},\gamma_{j,\tau},\psi_{j,\tau}^2)$ is a generalised inverse Gaussian distribution $\mathcal{GIG}(p,q,c)$, with $p=-0.5\rank(\mK_{j,\tau})+0.5$, $q=1/(r(\gamma_{j,\tau})\psi_{j,k}^2)$, $c=\betavec_{j,\tau}'\mK_{j,\tau}\betavec_{j,\tau}$. For Steps 3--5 it is easy to show that 
\[
\left. p(\gamma_{j,\tau}=1\,\right|\,\thetavec\backslash\gamma_{j,\tau}) = \left(1+\frac{\varphi(\zeta_{j,\tau};0; r_{j,\tau}\psi_{j,\tau}^2) (1-\omega_{j,\tau})}{\varphi(\tau_{j,\tau};0;\psi_{j,\tau}^2)\omega_{j,\tau}} \right)^{-1},
\]
 with $\varphi(\cdot; \mu, \sigma^2)$ the density of a normal distribution with mean $\mu$ and variance $\sigma^2$; and
\begin{equation*}\begin{aligned}
 \left.\psi_{j,\tau}^2\,\right|\, \thetavec\backslash\psi_{j,\tau}^2 &\sim \IGD\left(a_{j,\tau}+0.5,b_{j,\tau}+\frac{\tau_{j,\tau}^2}{2r_{j,\tau}(\gamma_{j,\tau})}\right)
\\
 \left.\omega_{j,\tau}\,\right|\,\thetavec\backslash\omega_{j,\tau} &\sim \BetaD(a_{0,j,\tau}+\gamma_{j,\tau},b_{0,j,\tau}+1-\gamma_{j,\tau}),
\end{aligned}\end{equation*}
To generate $w_i$, $i=1,\ldots,n$ at Step 6, we recall the i.i.d.~exponential priors with rate $\delta^2$ which implies full conditional distributions for the reciprocal weights which are inverse Gaussian:
\[
\left.w_i^{-1}\,\right|\,\thetavec\backslash w_i\sim\invGD\left(\sqrt{\frac{\xi^2+2\sigma^2}{(y_i-\eta_i)^2}},\frac{\delta^2(\xi^2+2\sigma^2)}{\sigma^2}\right).
\]
Finally, the full conditional distributions for $\delta^2$ at Step 7 are gamma, i.e.
\[
\left.\delta^2\,\right|\,\thetavec\backslash \delta^2\sim\GaD\left(a_\delta+\frac{3n}{2},b_\delta+\frac{1}{2\sigma^2}\sum_{i=1}^n\frac{(y_i-\eta_i-\xi w_i)^2}{w_i}+\sum_{i=1}^n w_i\right).
\]

\vspace{0.9em}
\paragraph{Implementation}
Our approach is  implemented in a developer version of the free software BayesX~\cite{BelBreKleKneLanUml2015Software}{, which} can be downloaded and compiled at \url{www.bayesx.org}. 
\section{Analysis on Madrid's air pollution}\label{sec:results}

While Sections~\ref{subsec:motivation} and~\ref{subsec:datades} provide a motivation for the problem underlying air pollution data and present the data and some initial exploratory analysis, respectively, this Section details our results. We recall Table \ref{tab:data} {shows} a description and summary statistics of the continuous covariates (before standardisation to $[0,1]$) and from which we can extract the full predictor specification
\begin{equation*}\begin{aligned}
\eta_{\tau}&=\sum_{k=0}^4 \beta_k + f_1(\mathit{co}) + f_2(\mathit{o3}) + f_3(\mathit{prec}) + f_4(\mathit{temp}) + f_5(\mathit{vel}) + f_6(\mathit{racha}) + f_7(\mathit{pres\_max})\\
&\quad + f_8(\mathit{pres\_min}) + f_9(\mathit{traffic})
\end{aligned}\end{equation*}
Here, $f_j=f_{j,\pen}+f_{j,\unpen}$ has been decomposed into respective linear and non-linear parts for each covariate (subject to selection) and $\beta_k$, $k=0,\ldots,3$ are the overall intercept and year-specific coefficients (not subject to selection). After inspection of Figures \ref{fig:prelim:no2:o3} and \ref{fig:prelim:no2:traffic}, we noted that NO2 distributions differ depending on the predictor values, and also show a non-Gaussian behaviour. We also underlined that the variance of NO2 increases with increasing traffic, and that the predictor O3 has a clear non-linear effect on NO2, while the one for traffic is rather linear.

\begin{table}[htbp]
\centering\renewcommand\arraystretch{1.00}
\begin{tabular}{c|cccc}
  \hline\hline
Covariate & $f_{\unpen}/f_{\pen}$ & $\tau=0.6$ & $\tau=0.8$ & $\tau=0.9$ \\ 
  \hline
{\emph{co}} & $f_{\unpen}$ (lin) &\textbf{ 1} & 0 & 0 \\ 
  {\emph{co}} & $f_{\pen}$ (non-lin) & 0.422 & 0.232 & 0.18 \\ 
  {\emph{o3}} & $f_{\unpen}$ (lin) & 0 & 0 & 0 \\ 
  {\emph{o3}} & $f_{\pen}$ (non-lin) & \textbf{1} & \textbf{0.997} & \textbf{1} \\ 
  {\emph{prec}} & $f_{\unpen}$ (lin) & 0 & 0 & 0 \\ 
  {\emph{prec}} & $f_{\pen}$ (non-lin) & \textbf{0.764} & \textbf{0.973} & \textbf{1} \\ 
  {\emph{temp}} & $f_{\unpen}$ (lin) & \textbf{1} & \textbf{1} & 0 \\ 
  {\emph{temp}} & $f_{\pen}$ (non-lin) &\textbf{ 0.958} &\textbf{ 0.992 }& \textbf{1 }\\ 
  {\emph{vel}} & $f_{\unpen}$ (lin) & \textbf{1} & 0 & 0 \\ 
  {\emph{vel}} & $f_{\pen}$ (non-lin) & 0.15 & \textbf{0.886} &\textbf{ 0.998 }\\ 
  {\emph{racha}} & $f_{\unpen}$ (lin) & 0 & \textbf{1} & 0 \\ 
  {\emph{racha}} & $f_{\pen}$ (non-lin) & \textbf{0.72}5 & \textbf{0.749} & \textbf{0.836} \\ 
  {\emph{pres\_max}} & $f_{\unpen}$ (lin) & 0 & 0 & 0 \\ 
  {\emph{pres\_max}} & $f_{\pen}$ (non-lin) & 0.36 & 0.111 & 0.353 \\ 
  {\emph{pres\_min}} & $f_{\unpen}$ (lin) & 0 & \textbf{1} & 0 \\ 
  {\emph{pres\_min}} & $f_{\pen}$ (non-lin) & 0.424 & 0.3 & 0.376 \\ 
  {\emph{traffic}} & $f_{\unpen}$ (lin) & 0 & \textbf{1} & 0 \\ 
  {\emph{traffic}} & $f_{\pen}$ (non-lin) & \textbf{0.945} & \textbf{0.813} & \textbf{0.865 }\\ 
   \hline\hline
\end{tabular}
\caption{\footnotesize Posterior mean inclusion probabilities of $f_{\unpen}$ (linear parts) and $f_{\pen}$ (non-linear parts) for $\tau\in\lbrace 0.6,0.8,0.9\rbrace$ (across columns 2--4). We say an effect part should be included in the model if the corresponding posterior mean inclusion probability $\dsP(\gamma_{j,\tau}|\thetavec\backslash\gamma_{j,\tau})\geq 0.5$.}\label{tab:inc}
\end{table}

\begin{figure}[htbp]
	\begin{center}
		\centering\includegraphics*[width=0.9\textwidth,angle=0]{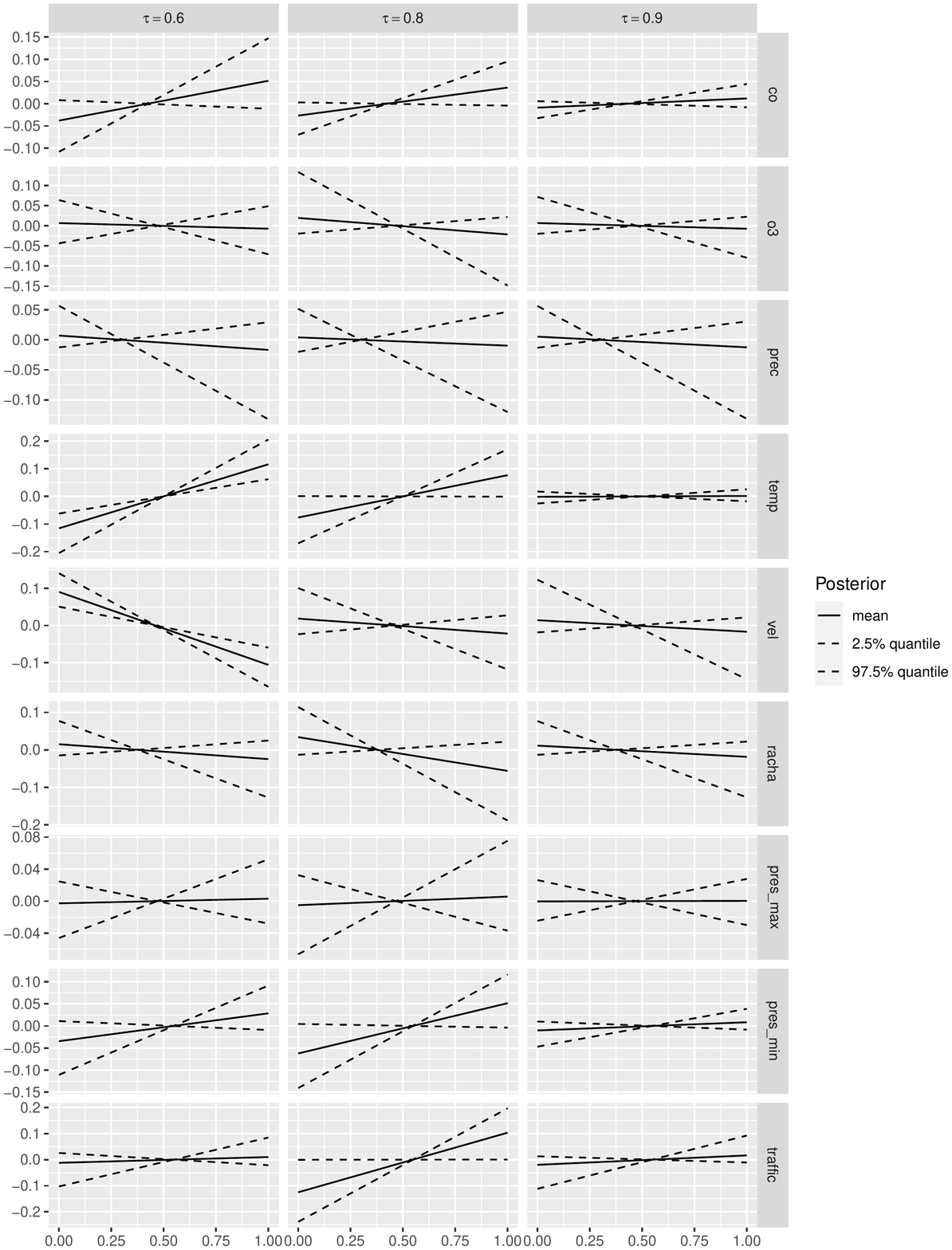}
	\end{center}
\caption{\footnotesize{Estimated posterior effects for $f_{\unpen}$ (linear parts) for $\tau\in\lbrace 0.6,0.8,0.9\rbrace$ (across columns 1--3) and for all nine covariates (row-wise). Shown are the posterior mean (solid lines) and 95\% pointwise credible intervals (dashed lines).}}
	\label{fig:no2_lin}
\end{figure}

\begin{figure}[htbp]
	\begin{center}
		\centering\includegraphics*[width=0.9\textwidth,angle=0]{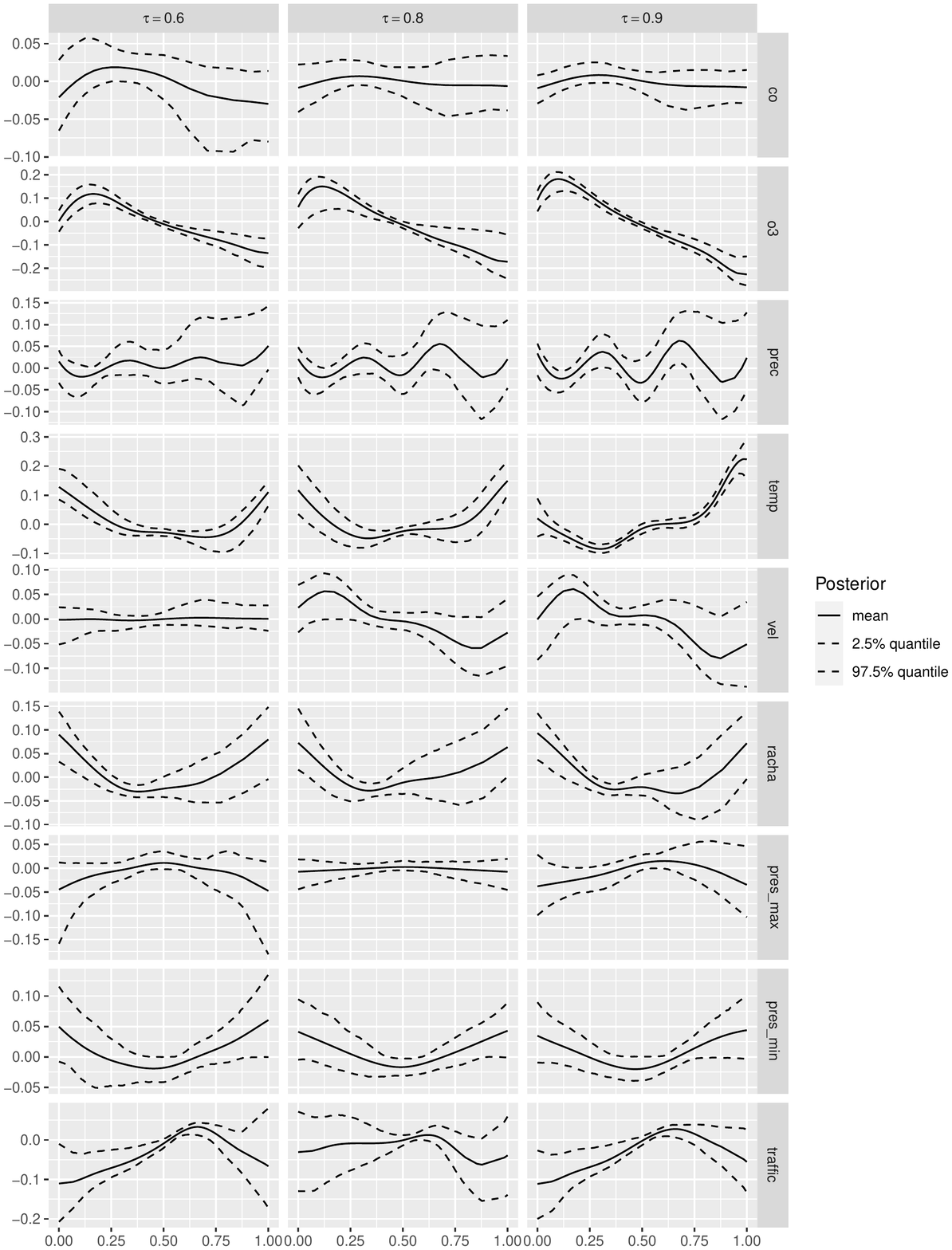}
	\end{center}
\caption{\footnotesize{Estimated posterior effects for $f_{\pen}$ (non-linear parts) for $\tau\in\lbrace 0.6,0.8,0.9\rbrace$ (across columns 1--3) and for all nine covariates (row-wise). Shown are the posterior mean (solid lines) and 95\% pointwise credible intervals (dashed lines).}}
	\label{fig:no2_nonlin}
\end{figure}

\begin{figure}[htbp]
	\begin{center}
		\centering\includegraphics*[width=0.9\textwidth,angle=0]{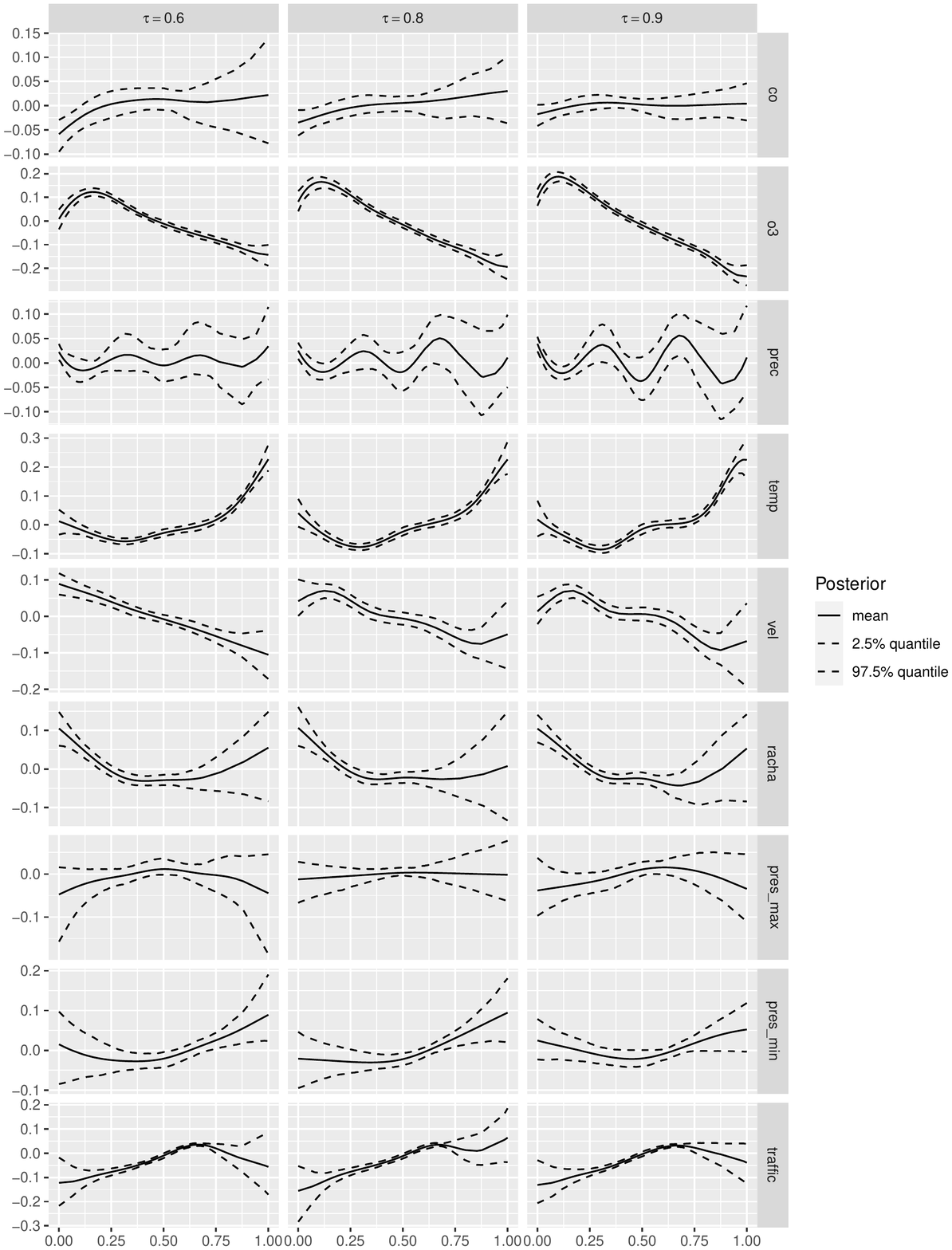}
	\end{center}
\caption{\footnotesize{Estimated posterior effects for $f=f_{\pen}+f_{\unpen}$ (linear parts+non-linear parts) for $\tau\in\lbrace 0.6,0.8,0.9\rbrace$ (across columns 1--3) and for all nine covariates (row-wise). Shown are the posterior mean (solid lines) and 95\% pointwise credible intervals (dashed lines).}}
	\label{fig:no2_both}
\end{figure}

In this line, we now discuss results for the conditional thresholds $\tau\in\lbrace 0.6,0.8,0.9\rbrace$. Table~\ref{tab:inc} shows posterior mean inclusion probabilities of $f_{\unpen}$ (linear parts) and $f_{\pen}$ (non-linear parts) for $\tau\in\lbrace 0.6,0.8,0.9\rbrace$ (across columns 2--4). We {say that} an effect part should be included in the model if the corresponding posterior mean inclusion probability $\dsP(\gamma_{j,\tau}|\thetavec\backslash\gamma_{j,\tau})\geq 0.5$ (in bold in Table~\ref{tab:inc}). In addition, Figures~\ref{fig:no2_lin} to~\ref{fig:no2_both} show estimated posterior effects for $f_{\unpen}$ (linear parts), $f_{\pen}$ (non-linear parts) and $f=f_{\pen}+f_{\unpen}$ (linear parts+non-linear parts) for $\tau\in\lbrace 0.6,0.8,0.9\rbrace$ (across columns 1--3) and for all nine covariates (row-wise). Shown are the posterior mean (solid lines) and 95\% pointwise credible intervals (dashed lines).

{We highlight the following results. CO ($\mathit{co}$) has a strong positive linear effect for $\tau=0.6$, but is negligible for $\tau=0.8,0.9$. This indicates that CO contributes to the NO2 distribution in a linear form but only for lower thresholds. In contract, O3 ($\mathit{o3}$) is a good predictor with a clear non-linear effect for all three quantiles with a notable decreasing effect. This inverse relationship between NO2 and O3 is expected as we discussed in Section \ref{subsec:motivation}. However, the non-linear structure was not so apparent, and we are able to underpin it. }

{Precipitation ($\mathit{prec}$) enters as a non-linear effect, and varies its strength depending on the quantile. In particular, the effect and its non-linearity is increasing with raising quantile $\tau$. This combination of non-linear behaviour and increasing strength with raising quantiles brings a clear explanation of the effect of precipitation over NO2. Average temperature ($\mathit{temp}$) enters as both  non-linear and linear effects for $\tau=0.6,0.9$  but only as a non-linear effect for $\tau=0.8,0.9$. Thus, the quantile has an effect on the dependence between  NO2 and average temperature with the higher the temperature, the larger NO2.}

Average wind speed ($\mathit{vel}$) enters as a non-linear effect for $\tau=0.8,0.9$ and linearly for $\tau=0.6$. The linear effect for small thresholds is inverse indicating that with an increasing wind speed we get decreasing NO2 values. Importantly, for larger thresholds this linearity vanishes towards non-linear effects. When coming to maximum wind gusts ($\mathit{racha}$), the effects are basically  non-linear for all quantiles. Air pressure ($\mathit{pres\_min}$) is only relevant as a linear effect for $\tau=0.8$ measured by minimum air pressure, but negligible for all other quantiles. Maximum pressure ($\mathit{pres\_max}$) is however not selected.

Finally, average traffic flow ($\mathit{traffic}$)  enters as a non-linear effect, with a stronger effect for $\tau=0.6,0.9$ and weaker effect for $\tau=0.8$. This indicates that strong traffic congestion\ns{s} affect NO2 in a complicated non-linear fashion, but also this holds for lower thresholds, probably due to  cross-relationships amongst some of the covariates. Latent (unobserved variables) also might place a hidden effect here, difficult to account for.

Noting that modelling reality of air pollution is 
certainly a critical, while complicated environmental problem, we have detected some functional forms, some of them highly non-linear, that underline the cross-relationships between a number of covariates and NO2. Chemical reactions are playing a role and make things even more complicated. Our statistical approach has been able to clarify some of these complicated relations.

\section{Discussion and conclusions}\label{sec:conclusion}

This work fills the existing gap of methods for effect selection in semi-parametric quantile regression models. Inspired by the recent work of \citet{KleCarKneLanWag2021} in the context of structured additive distributional regression~\citep{KleKneKlaLan2015} our Bayesian effect selection approach for additive quantile regression employs a normal beta prime {spike and slab prior} on the scalar squared importance parameters associated with each effect part in the predictor. Compared to the distributional models of~\cite{KleKneKlaLan2015} where predictors are placed on the distributional parameters, our quantile regression is better interpretable as it allows to directly select certain effect types on conditional quantiles of a response  and to decide whether relevant predictors affect the quantiles linearly or non-linearly. Other than in  structured additive {distributional} regression, MCMC is extremely fast in our approach since all steps can be realised in Gibbs updates. Furthermore, we solve the large computational burden for eliciting the prior hyperparameters by making use of the scaling property of the (scaled) beta prime distribution. 

While our method can be useful in a wide range of applications when interest is in understanding the impacts of influential covariates on the conditional distribution of a dependent variable beyond the mean, our methodological developments have specifically been stimulated the largest European environmental health risk, namely air pollution. Many European cities regularly exceed NO2 limits 
and we consider one weather and pollution station in downtown of Madrid as a representative.  We believe that we have better approximated the reality of air pollution  detecting complicated linear and non-linear functional forms in combination with particular alarm thresholds.  The results of this study are easily applicable to many other cities worldwide, perhaps with some adaptation and use of additional covariates. Indeed, our  statistical approach enables to study  quantile-specific covariate effects of any general functional form, and it helps deciding whether an effect should be included linearly, non-linearly or not at all in the relevant threshold quantiles.  

We should note at this point that this paper only considers a representative station in a big city such as Madrid. The reason for this is that the focus of this paper is the analysis of the inherent relationships between a number of predictors and NO2 to better understand the intrinsic mechanisms underlying air pollution, and the focus is not on prediction. These complicated mechanisms are missed out or simply not able to be understood by using other more widely encountered statistical approaches.  We are also aware that there are more measuring stations spread through the city and the spatial structure could be relevant if the focus would be more on prediction of missing data, or prediction onto the future. We leave this important point for future extensions of our approach. 

\newpage

\singlespacing
\renewcommand{\baselinestretch}{1.0}

\bibliographystyle{dcu}
\bibliography{references}
\end{document}